\documentclass[10pt,british,english,aps,superscriptaddress,preprintnumbers,amsmath,nofootinbib,amssymb,altaffilletter]{revtex4}
\usepackage[T1]{fontenc}
\usepackage[latin9]{inputenc}
\usepackage{color}
\usepackage{babel}
\usepackage{float}
\usepackage{amsmath}
\usepackage{amssymb}
\usepackage{graphicx}
\usepackage{wasysym}
\usepackage{esint}
\usepackage[unicode=true,pdfusetitle,
 bookmarks=true,bookmarksnumbered=false,bookmarksopen=false,
 breaklinks=false,pdfborder={0 0 1},backref=false,colorlinks=false]
 {hyperref}

\makeatletter

\providecommand{\tabularnewline}{\\}

\@ifundefined{textcolor}{}
{%
 \definecolor{BLACK}{gray}{0}
 \definecolor{WHITE}{gray}{1}
 \definecolor{RED}{rgb}{1,0,0}
 \definecolor{GREEN}{rgb}{0,1,0}
 \definecolor{BLUE}{rgb}{0,0,1}
 \definecolor{CYAN}{cmyk}{1,0,0,0}
 \definecolor{MAGENTA}{cmyk}{0,1,0,0}
 \definecolor{YELLOW}{cmyk}{0,0,1,0}
}

\usepackage{babel}
\usepackage{babel}
\usepackage{babel}
\usepackage{babel}

\usepackage{babel}
\usepackage{breakurl}

\newcommand{\LF}{\left(}
\newcommand{\RF}{\right)}
\newcommand{\LT}{\left[}
\newcommand{\RT}{\right]}

\@ifundefined{textcolor}{}{%
 \definecolor{BLACK}{gray}{0}
 \definecolor{WHITE}{gray}{1}
 \definecolor{RED}{rgb}{1,0,0}
 \definecolor{GREEN}{rgb}{0,1,0}
 \definecolor{BLUE}{rgb}{0,0,1}
 \definecolor{CYAN}{cmyk}{1,0,0,0}
 \definecolor{MAGENTA}{cmyk}{0,1,0,0}
 \definecolor{YELLOW}{cmyk}{0,0,1,0}
}

\usepackage{babel}
\usepackage{epsfig}\usepackage{mathrsfs}\usepackage{euscript}

\renewcommand{\[}{\begin{equation}}
\renewcommand{\]}{\end{equation}}

\makeatother

\begin{document}

\title{Conformal GUT inflation, proton lifetime and non-thermal leptogenesis}

\author{K. Sravan Kumar}
\email{sravan@ubi.pt, korumilli@sustc.edu.cn, sravan.korumilli@rug.nl}

\selectlanguage{british}%

\address{Departamento de F\'{i}sica, Centro de Matem\'atica e Aplica\c{c}\~oes
(CMA-UBI), Universidade da Beira Interior, 6200 Covilh\~a, Portugal}

\address{Southern University of Science and Technology (SUSTech), Shenzhen, 518055, China}
\selectlanguage{english}%

\address{Van  Swinderen  Institute,  University  of  Groningen,  9747  AG,  Groningen,  The  Netherlands}



\author{Paulo Vargas Moniz}
\email{pmoniz@ubi.pt}

\selectlanguage{british}%

\address{Departamento de F\'{i}sica, Centro de Matem\'atica e Aplica\c{c}\~oes
(CMA-UBI), Universidade da Beira Interior, 6200 Covilh\~a, Portugal}
\selectlanguage{english}%

\address{DAMTP, Wilberforce Road, University of Cambridge CB3 0WA, United Kingdom}

\address{Clare Hall, Hershel Road, University of Cambridge CB3 9AL, United Kingdom}

\begin{abstract}
In this paper, we generalize Coleman-Weinberg (CW) inflation in grand
unified theories (GUTs) such as $\text{SU}(5)$ and $\text{SO}(10)$
by means of considering two complex singlet fields with conformal
invariance. In this framework, inflation emerges from a spontaneously
broken conformal symmetry. The GUT symmetry
implies a potential with a CW form, as a consequence of radiative corrections.
The conformal symmetry flattens the above VEV branch of the CW potential to a Starobinsky plateau. As a result, we obtain $n_{s}\sim 1-\frac{2}{N}$
and $r\sim \frac{12}{N^2}$ for $N\sim 50-60$ e-foldings. Furthermore, this framework allow us to  estimate the proton lifetime as $\tau_{p}\lesssim 10^{40}$
years, whose decay is mediated by the superheavy gauge bosons. Moreover, we implement
a type I seesaw mechanism by weakly coupling the complex singlet, which
carries two units of lepton number, to the three generations of singlet
right handed neutrinos (RHNs). The spontaneous symmetry breaking
of global lepton number amounts to the generation of neutrino masses.
We also consider non-thermal leptogenesis in which the inflaton dominantly
decays into heavy RHNs that sources the observed baryon asymmetry.
We constrain the couplings of the inflaton field to the RHNs,
which gives the reheating temperature as $10^{6}\text{ GeV}\lesssim T_{R}<10^{9}$
GeV. 
\end{abstract}

\date{\today}
\maketitle

\section{Introduction}

Primordial inflation is a successful paradigm for the description
of the early Universe and it is strongly supported by the current
observational data \cite{Starobinsky:1980te,Guth:1980zm,Linde:1981mu,Ade:2015lrj}. Primordial perturbations,
when the scales exiting the horizon $\left(k\sim aH\right)$, are
eventually responsible for the structure formation in the Universe.
From {\it Planck} 2015 \cite{Ade:2015lrj,Ade:2015tva}, the key observables
of inflation, namely, the scalar tilt and the ratio of tensor to scalar
power spectra, are constrained as $n_{s}=0.968\pm0.006$, $r<0.09$
at $95\%$ confidence level. The CMB power spectra is observed to
be nearly adiabatic, scale invariant and Gaussian \cite{Ade:2015ava,Ade:2015lrj}.
Although the physical nature of the inflaton is still uncertain \cite{Martin:2013tda,Martin:2015dha},
the models based on $f(R)$ or a canonical scalar field with a flat potential are
favoured with respect to the data. Since the inflationary scale
is in general expected to be $\sim10^{16}\,\text{GeV}$, it is natural
consider the inflaton to be a scalar field associated with grand unified
theory (GUT) groups, such as $\text{SU}(5)$ and $\text{SO}(10)$.
The Shafi-Vilenkin (SV) model \cite{Shafi:1983bd} is one of the first
realistic model of inflation which was based on $\text{SU}(5)$ GUT
\cite{Georgi:1974sy}. In this framework, inflation is a result of the
spontaneous breaking of $\text{SU}(5)\to\textrm{SU}(3)_{c}\times\textrm{SU}(2)_{L}\times\textrm{U}(1)_{Y}$
by a GUT field ($\textbf{24-\text{plet}}$ adjoint Higgs) and an inflaton,
which is a SU(5) singlet that rolls down to a vacuum expectation value
(VEV). The success of the SV model is that it can lead to a successful
baryogenesis after inflation and predicts a proton life time above
the current lower bound \cite{Shafi:2006cs,Rehman:2008qs}. In this model,
the scalar field potential is of a Coleman-Weinberg (CW) form,
according to which primordial gravitational waves are constrained
by $0.02\le r\le0.1$ \cite{Okada:2014lxa}. Although the SV model is
well within the current bounds of {\it Planck} 2015, several extensions of this
model were studied to get smaller values of $r$. In \cite{Cerioni:2009kn,Panotopoulos:2014hwa,Barenboim:2013wra,Kannike:2015kda,Racioppi:2017spw},
CW inflation was studied in the context of induced gravity, non-minimal
coupling and brane-world scenario, where the tensor to scalar ratio
was obtained to be $r\sim\mathcal{O}\left(10^{-2}\right)-\mathcal{O}\left(10^{-3}\right)$.
We note that all these modifications necessarily introduce an additional
parameter whose value determines the shape of the inflaton potential in the Einstein frame\footnote{For example, in the case of non-minimal coupling of inflaton to Ricci scalar ($\xi$), the value of $\xi\gg 1$ in order to get shape of the potential to be like a Starobinsky plateau \cite{Kallosh:2013tua,Broy:2016rfg}. }. 

Moreover, extensions of the SV model within particle physics offer
rich physics beyond the Standard Model (SM). Therefore, the SV model is embedded
in a higher gauge group such as $\text{SO}\left(10\right)$, which can
be broken to the SM via an intermediate group $\text{G}_{422}=\text{SU}(4)_{c}\times\text{SU}\left(2\right)_{L}\times\text{SU}\left(2\right)_{R}$
\cite{Lazarides:1984pq,Lazarides:1991wu}. Obtaining successful inflation
in $\text{SO}\left(10\right)$ is more realistic with additional benefits
to explain physics beyond SM, such as neutrino physics, matter anti-matter
asymmetry through non-thermal leptogenesis, monopoles and dark matter
(DM) \cite{Rehman:2008qs}. For example, Ref. \cite{Boucenna:2014uma}
considered a complex singlet scalar being coupled to right handed neutrinos (RHNs), followed by implementing type I seesaw mechanism. This
approach unified inflation with Majorana DM together with the scheme
of generating neutrino masses. In \cite{Okada:2013vxa} an additional
$\text{U}(1)_{B-L}$ symmetry was considered in the SM i.e., $\textrm{SU}(3)_{c}\times\textrm{SU}(2)_{L}\times\textrm{U}(1)_{Y}\times\text{U}\left(1\right)_{B-L}$,
where $B-L$ symmetry can be spontaneously broken when the scalar
field takes the VEV. In this setup, we can explain the baryon asymmetry
of the Universe through non-thermal leptogenesis \cite{Lazarides:1991wu,Asaka:2002zu,Senoguz:2003hc,Senoguz:2005bc}.
Recently, CW inflation was studied in an extension with $\text{SO}(10)$
and $\text{E}_{6}$, pointing out the possibility of observing primordial
monopoles \cite{Senoguz:2015lba}. 

Apart from models based on GUT theories, the Starobinsky model based
on the $R^{2}$ gravity modification and the Higgs inflation \cite{Starobinsky:1980te,Starobinsky:1983zz,Bezrukov:2007ep}
occupy a privileged position, with practically equal predictions in
the $\left(n_{s},\,r\right)$ plane 
\begin{equation}
n_{s}=1-\frac{2}{N}\quad,\quad r=\frac{12}{N^{2}}\,,\label{targetspot}
\end{equation}
where $N$ is the number of e-foldings before the end of inflation.
There has been a growing interest on embedding these models in string
theory and supergravity (SUGRA) aiming for a UV completion \cite{Linde:2014nna,Ferrara:2013rsa}.
Recently, a UV completion of the Starobinsky model was proposed in the context
of non-local gravity inspired from string field theory \cite{Koshelev:2016xqb,Koshelev:2017tvv}.
Starobinsky like models were also developed in $\mathcal{N}=1$ SUGRA,
namely, no scale \cite{Ellis:2013nxa} and $\alpha-$attractor models
\cite{Kallosh:2013yoa} where an additional physical parameter leads to any value of $r<0.1$. In \cite{Kumar:2015mfa} $\alpha-$attractor
models were studied in the non-slow-roll context where a new class
of potentials were shown to give the same predictions. On the other hand,
Higgs inflation is particularly interesting due to the fact that Higgs
was the only scalar so far found at LHC. But for it to be an inflaton candidate compatible with CMB data, we require a very large non-minimal coupling $\left(\xi\gg 1\right)$
to Ricci scalar. It was known that a scalar field with large non-minimal
coupling gives rise to a $R^{2}$ term considering 1-loop quantum corrections.
Consequently, renormalization group (RG) analysis shows that Higgs
inflation is less preferable compared to Starobinsky model \cite{Salvio:2015kka,Calmet:2016fsr}.
This result not only applies to Higgs inflation but also to any
arbitrary scalar with very large non-minimal coupling. Furthermore, in
both $R^{2}$ and Higgs inflation the inflaton field rolls down to
zero after inflation\footnote{Although the SM Higgs field rolls to its electroweak VEV it is negligible compared to
the energy scale of inflation.}. Differently, in GUT theories the inflaton field acquires a VEV
due to its interaction with GUT fields. 

The main goal of this paper is to generalize the SV model in order to
achieve $r\sim\mathcal{O}\left(10^{-3}\right)$ without introducing
any additional parameters that affect the flatness of the inflaton potential (in Einstein frame), coasting towards a Starobinsky plateau\footnote{{Our construction provides an alternative way from the scalar field models with large non-minimal coupling  $\xi$ \cite{Kallosh:2013tua,Broy:2016rfg}}}.
In our construction, we introduce conformal symmetry (or local scale
invariance) in a GUT model. It was shown by Wetterich \cite{Wetterich:1987fm}
that scale symmetries play a crucial role in the construction of realistic
cosmological models based on particle physics. Moreover, scale symmetries
successfully explain the hierarchy of different scales such as the Planck
and the Higgs mass \cite{Hooft:2014daa,Quiros:2014hua,Scholz:2012ev,Bars:2013yba}.
Therefore, it is natural to consider scale invariance in constructing an
inflationary scenario, through which we can obtain a dynamical generation of the
Planck mass, inflationary scale and particle physics scales beyond SM. In this regard, we
consider two complex singlet fields $\left(\bar{X},\,\Phi\right)$
of $\text{SU}(5)$ or $\text{SO}(10)$ and couple them to the Ricci scalar
and adjoint Higgs field $\left(\Sigma\right)$, such that the total
action would be conformally invariant. We obtain inflation as a result
of spontaneous breaking the conformal and GUT symmetries. The former
occurs due to gauge fixing of one singlet field to a constant for all spacetime and the
latter occurs due to $\Sigma$ field taking its GUT VEV. Here the inflaton
is identified with the real part of the second singlet ($\phi=\sqrt{2}\mathfrak{Re}\left[\Phi\right]$),
whereas the imaginary part is the corresponding Nambu-Goldstone boson
is assumed to pick up a mass due to the presence of small explicit
soft lepton number violation terms in the scalar potential \cite{Boucenna:2014uma}.
We also assume $\Phi$ carries two units of lepton number and it
is coupled to the RHNs.
Near the end of inflation, the inflaton is supposed to reach its VEV and
also the global lepton number is violated. Thereafter, we study the
dominant decay of inflaton into heavy RHNs producing non-thermal
leptogenesis. We compute the corresponding reheating temperatures
and also discuss the issue of producing the observed baryon asymmetry.
Our study completes with an observationally viable inflationary
scenario, predicting proton life time, neutrino masses and producing
non-thermal leptogenesis from heavy RHNs. 

The paper is briefly organized as follows. In Sec. \ref{SCmodel-sec},
we describe toy models with conformal and scale invariance. We identify
the interesting aspects of spontaneous symmetry breaking of these symmetries leading to
viable inflationary scenarios. In Sec. \ref{SU5CW-sec}, we briefly
present the SV model and the computation of the proton
life time. In Sec. \ref{twofieldmodel-sec} we propose our generalization
of the SV model by introducing an additional conformal symmetry. We report
the inflationary predictions of the generalized model together with estimates
of proton life time. In Sec. \ref{seesawSec} we further explore the
nature of inflaton couplings to the SM Higgs and singlet RHNs
through type I seesaw mechanism. In the view of the dominant decay of the inflaton into heavy RHNs, we constrain the Yukawa couplings
of the inflaton field compatible with the generation of light left handed neutrino
masses. In Sec. \ref{ReheatSec} we implement non-thermal leptogenesis
and compute the reheating temperatures corresponding to the dominant
decay of inflaton to heavy RHNs. We additionally comment on
the necessary requirements for the production of observed baryon asymmetry
through CP violation decays of RHNs. In Sec. \ref{conclusions}
we summarize our results pointing to future steps. We provide an Appendix.~\ref{A1} summarizing the effects of geometric destabilization from fields space of inflaton and the presence of heavy fields in our model. 
In this paper we follow the units $\hbar=1,\,c=1,\,$ $m_{\rm P}^{2}=\frac{1}{8\pi G}$.

\section{Conformal vs Scale invariance}

\label{SCmodel-sec}

Models with global and local scale invariance (Weyl invariance (or)
conformal invariance) are often very useful to address the issue
of hierarchies in both particle physics and
cosmology \cite{Englert:1976ep,Deser:1970hs,Hooft:2014daa,Shaposhnikov:2008xi,Scholz:2012ev,Quiros:2014hua}.
Models with these symmetries contains no input mass parameters.
The spontaneous breaking of those symmetries induced by the VEV's
of the scalar fields present in the theory, generates a hierarchy of
mass scales e.g., Planck mass, GUT scale and neutrino masses\footnote{For example, single scalar field models with spontaneously broken scale invariance by
the 1-loop corrections 
were studied in
\cite{Rinaldi:2015uvu,Rinaldi:2015yoa,Csaki:2014bua}. In \cite{Ferreira:2016vsc} a
two field model with scale invariance was studied to
generate the hierarchy of mass scales and the dynamical generation of
Planck mass from the VEV's of the scalar fields. Recently in \cite{Kannike:2016wuy},
some constraints were derived on these models from Big Bang Nucleosynthesis
(BBN).}. Moreover, it is a generic feature that scale or conformal symmetry breaking
induce a flat direction in the scalar field potential \cite{Wetterich:1987fm},
which makes these models even more interesting in the context of inflation.
Another motivation to consider scale invariance for inflationary model
building comes from CMB power spectra which is found to be nearly
scale invariant \cite{Ade:2015lrj}.

In this section, we present firstly a toy model (with two fields) that
is (global) scale invariant and present the generic form of (scale invariant)
potentials and their properties. We review the presence of a massless
Goldstone boson that appears as a result of spontaneous breaking of global scale invariance. In the following, we discuss the two field conformally invariant
model, in which case the presence of a massless Goldstone boson can be removed
by appropriate gauge fixing. The resultant spontaneous breaking
of conformal symmetry (SBCS) turns to be very useful to obtain a Starobinsky
like inflation\footnote{The role of SBCS was discussed in Higgs-dilaton models of inflation and dark energy \cite{GarciaBellido:2011de,Bezrukov:2012hx,Karananas:2016kyt,Rubio:2017gty}}\footnote{Toy models of conformal inflation were studied in \cite{Kallosh:2013lkr,Kallosh:2013daa}
and were embedded in $\mathcal{N}=1$ SUGRA. Furthermore, in a recent study
conformal models were shown to be motivated in the context of string
field theory \cite{Koshelev:2016vhi}.}. We will later explore the role of SBCS in a more realistic
inflationary setting based on GUTs. 

\subsection{Scale invariance }

\label{sec-SI}

Here we discuss a toy model with two scalar fields (in view of Refs.
\cite{Wetterich:1987fk,Wetterich:1987fm,Ghilencea:2015mza,Ferreira:2016vsc})
and point out interesting features that we later utilize
in our construction.

A generic two field global scale invariant action can be written as

\begin{equation}
S_{global}=\int d^{4}x\,\sqrt{-g}\left[\frac{\alpha}{12}\phi^{2}R+\frac{\beta}{12}\chi^{2}R-\frac{1}{2}\partial^{\mu}\phi\partial_{\mu}\phi-\frac{1}{2}\partial^{\mu}\chi\partial_{\mu}\chi-\phi^{4}f\left(\rho\right)\right]\,,\label{SIaction}
\end{equation}
where $\alpha,\,\beta$ are constants and $\rho=\frac{\phi}{\chi}$,
the generic function $f\left(\frac{\phi}{\chi}\right)$ here can be
treated as quartic self coupling of the field $\phi$ \cite{Wetterich:1987fm,Ghilencea:2015mza}.
The action (\ref{SIaction}) is scale invariant, i.e., invariant under
global scale transformations $g_{\mu\nu}\to e^{-2\lambda}g_{\mu\nu}\,,\,\phi\to e^{\lambda}\phi\,,\,\chi\to e^{\lambda}\chi$
for any constant $\lambda$ (dilatation symmetry). 

Since the potential $V\left(\phi,\,\chi\right)=\phi^{4}f\left(\rho\right)$
is homogeneous, it must satisfy the following constraint \cite{Ghilencea:2015mza,Ferreira:2016vsc}

\begin{equation}
\phi\frac{\partial V}{\partial\phi}+\chi\frac{\partial V}{\partial\chi}=4V\,.\label{homo-cd}
\end{equation}
The extremum conditions for $V$, i.e., $\partial_{\phi}V=\partial_{\chi}V=0$
can also be written as $f\left(\rho\right)=f^{\prime}\left(\rho\right)=0$.
One of the conditions fix the ratio of the VEV's of the fields, while the
other gives a relation between couplings (if $\langle\phi\rangle\neq0$
and $\langle\chi\rangle\neq0$). The interesting property 
here is that if $\langle\phi\rangle\propto\langle\chi\rangle$ there
exists a flat direction for the field $\phi$ (see \cite{Wetterich:1987fm}
for detailed analysis). This will be more useful in the context of local scale invariant model.

Let us consider a scale invariant potential of the form

\begin{equation}
V_{1}=\frac{\lambda_{\phi}}{4}\phi^{4}+\frac{\lambda_{m}}{2}\phi^{2}\chi^{2}+\frac{\lambda_{\chi}}{4}\chi^{4}\,,\label{pot-coupling}
\end{equation}
where the couplings can in general depend on the ratio of the two fields
i.e., $\phi/\chi$. If for example, we assume the couplings to be independent
of the ratio of the two fields and consider the spontaneous breaking of
scale symmetry i.e., the case with $\langle\phi\rangle\neq0,\,\langle\chi\rangle\neq0$,
thus, as a result of minimizing the potential, we arrive at \cite{Ghilencea:2015mza}

\begin{equation}
\frac{\langle\phi\rangle}{\langle\chi\rangle}=-\frac{\lambda_{m}}{\lambda_{\phi}}\quad,\quad V_1=\frac{\lambda_{\chi}}{4}\left(\chi^{2}+\frac{\lambda_{m}}{\lambda_{\chi}}\phi^{2}\right)^{2}\,,\label{pot-SI1}
\end{equation}
with $\lambda_{m}^{2}=\lambda_{\phi}\lambda_{\chi}$ and $\lambda_{m}<0$.

In (\ref{pot-SI1}) we can re-define the coupling as

\begin{equation}
\bar{\lambda}_{\chi}=\lambda_{\chi}\left(1+\frac{\lambda_{m}}{\lambda_{\chi}}\frac{\phi^{2}}{\chi^{2}}\right)^{2}\,,\label{coupling-SI}
\end{equation}
then the potential (\ref{pot-SI1}) looks like a simple quartic potential

\begin{equation}
V_{1}=\frac{\bar{\lambda}_{\chi}}{4}\chi^{4}\,.\label{pot-SI2}
\end{equation}
We can alternatively have a potential of the form 

\begin{equation}
V_{2}=\frac{\tilde{\lambda}_{\phi}}{4}\phi^{4}\,,\quad\tilde{\lambda}_{\phi}=\lambda_{\phi}\left(1-\frac{\phi^{2}}{\chi^{2}}\right)^{2}\,,\label{pot-real}
\end{equation}
which also satisfies the constraint (\ref{homo-cd}) and is 
different from (\ref{pot-coupling}). We will later see that the form
of the potential in (\ref{pot-real}) gives a viable inflationary scenario. From (\ref{pot-SI1}) -(\ref{pot-real}) we can crucially
learn how to define couplings as a function of the ratio of
two fields in a scale invariant model. Of course, we only considered
here simple toy models. However, we note that such field dependent
couplings can be expected to arise in string theory and were applied
in the context of early Universe \cite{Kao:1990tp}. 

The spontaneous breaking of scale symmetry occurs when one of the fields develops a VEV (let us take the field $\chi$)
which induces the emergence of a corresponding massless Goldstone
boson (dilaton) $\tilde{\chi}=\sqrt{6}M\ln\left(\frac{\chi}{\sqrt{6}M}\right)$
that is associated with an arbitrary scale $M\propto m_{\rm P}$
\cite{Wetterich:1987fm}. By performing a Weyl rescaling of the metric
$g_{\mu\nu}\to\tilde{g}_{\mu\nu}=\left(\frac{\chi}{\sqrt{6}M}\right)^{2}g_{\mu\nu}$
and $\phi\to\tilde{\phi}=\frac{M}{\sqrt{6}\chi}\phi$ we indeed observe
that the field $\tilde{\chi}$ is massless since the
potential becomes independent of the field $\tilde{\chi}$

\begin{equation}
V\left(\phi,\,\chi\right)=\phi^{4}f\left(\frac{\phi}{\chi}\right)=\tilde{\phi}^{4}f\left(\frac{\tilde{\phi}}{M}\right)\,.\label{pot-SIB}
\end{equation}
Although interesting cosmology and particle physics can be developed
based on the scale invariant models, we need to constrain the
implications of the massless dilaton present in the system \cite{Bars:2013yba}.
It was shown that the dilaton can be gauged away if we consider a
model with local scale symmetry \cite{Bars:2012mt}. 

\subsection{Conformal invariance }

\label{CS}

A general action that is invariant under local scale transformations
$g_{\mu\nu}\to\Omega^{-2}\left(x\right)g_{\mu\nu}\,,\,\phi\to\Omega(x)\phi\,,\,\chi\to\Omega(x)\chi$
can be written as 

\begin{equation}
S_{local}=\int d^{4}x\,\sqrt{-g}\left[\frac{\left(\chi^{2}-\phi^{2}\right)}{12}R+\frac{1}{2}\partial^{\mu}\chi\partial_{\mu}\chi-\frac{1}{2}\partial^{\mu}\phi\partial_{\mu}\phi-\phi^{4}f\left(\frac{\phi}{\chi}\right)\right]\,,\label{toy}
\end{equation}
where the potential in the above action should also satisfy the condition (\ref{homo-cd}).

From the above action we can define an effective Planck mass $m_{eff}^{2}=\frac{\chi^{2}-\phi^{2}}{6}$
which evolves with time. In these theories, we would recover the standard
Planck scale $m_{\rm P}$ when the fields reach their VEV. Note that the
field $\chi$ contains a wrong sign for the kinetic term but it is not a
problem as we can gauge fix the field at $\chi=\text{constant}=\sqrt{6}M$
for all spacetime where $M\sim\mathcal{O}\left(m_{\rm P}\right)$. This
particular gauge choice is so called $c-$gauge\footnote{Supergravity gauge was first realized in the context of $2T-$ physics based SUGRA models \cite{Bars:2012mt,Bars:2013yba} where it was shown to be useful to obtain geodesic completeness of the theory. {We follow this gauge choice in this paper as it allow us to explain hierarchy of scales in our model.}} which spontaneously breaks
the conformal symmetry. It was argued that the theories in this
gauge are of interest especially in cosmological models based on particle
physics \cite{Bars:2013yba}. {We will further see in this paper that fixing the scale $M$ sources the hierarchy of mass scales related to inflation and particle physics (e.g., neutrino masses).} In the inflationary models based on GUTs it is
natural that the field $\phi$ takes a non-zero VEV, i.e., $\langle\phi\rangle\neq0$
in which case it is useful to assume $6M^{2}-\langle\phi\rangle^{2}=6m_{\rm P}^{2}$
in order to generate Planck mass. Moreover, its also necessary to
keep the evolution of the field $\phi\lesssim\sqrt{6}M$ in order
to avoid an anti-gravity regime.

Considering $f\left(\frac{\phi}{\chi}\right)=\lambda\left(1-\frac{\phi^{2}}{\chi^{2}}\right)^{2}$
in (\ref{toy}), SBCS via gauge fixing $\chi=\sqrt{6}m_{\rm P}$ leads
to the Einstein frame action in terms of a canonically normalized field
$\phi=\sqrt{6}m_{\rm P}\,\tanh\left(\frac{\varphi}{\sqrt{6}m_{\rm P}}\right)$
and it is written as

\begin{equation}
S_{local}=\int d^{4}x\,\sqrt{-g}\left[\frac{m_{\rm P}^{2}}{2}R-\frac{1}{2}\partial^{\mu}\varphi\partial_{\mu}\varphi-\lambda m_{\rm P}^{4}\tanh^{4}\left(\frac{\varphi}{\sqrt{6}m_{\rm P}}\right)\right]\,.\label{toy-1}
\end{equation}
We can see that the above action leads to a Starobinsky like inflation
as the potential acquires a plateau when $\varphi\gg m_{\rm P}$ (i.e., $\phi\to\sqrt{6}m_{\rm P}$).
In this case the inflaton rolls down to a zero VEV by the end of inflation because of the gauge fixing $\chi=\sqrt{6}m_{\rm P}$ and consequently Einstein gravity is recovered.

In the next sections, we will study realistic GUT inflationary models where the inflaton
rolls down to non-zero VEV and sources interesting implications in particle physics sector. 

\section{Coleman-Weinberg GUT inflation}

\label{SU5CW-sec}

In this section we briefly review the Shafi-Vilenkin model \cite{Shafi:1983bd,Linde:2005ht}.
It is one of the first realistic model of inflation which was based
on SU(5) grand unified theory (GUT) . In this framework a new scalar
field $\phi$, a SU(5) singlet was considered and it weakly interacts
with the GUT symmetry breaking field (adjoint) $\Sigma$ and fundamental
Higgs field $H_{5}$. The tree level scalar potential is given by

\begin{equation}
\begin{aligned}V\left(\phi,\,\Sigma,\,H_{5}\right)= & \frac{1}{4}a\left(\textrm{Tr}\Sigma^{2}\right)^{2}+\frac{1}{2}b\textrm{Tr}\Sigma^{4}-\alpha\left(H_{5}^{\dagger}H_{5}\right)\textrm{Tr}\Sigma^{2}+\frac{\beta}{4}\left(H_{5}^{\dagger}H_{5}\right)^{2}\\
 & +\gamma H_{5}^{\dagger}\Sigma^{2}H_{5}+\frac{\lambda_{1}}{4}\phi^{4}-\frac{\lambda_{2}}{2}\phi^{2}\textrm{Tr}\Sigma^{2}+\frac{\lambda_{3}}{2}\phi^{2}H_{5}^{\dagger}H_{5}\,.
\end{aligned}
\label{TreeSU5}
\end{equation}
where the coefficients $a,\,b,\,\alpha$ and $\beta$ are taken to
be of the order of\footnote{The field $\Sigma$ interacts with vector boson $X$ with a coupling
constant $g$} $g^{2}$, therefore the radiative corrections in $\left(\Sigma,\,H_{5}\right)$
sector can be neglected. The coefficient $\gamma$ takes a relatively
smaller value and $0<\lambda_{i}\ll g^{2}$ and $\lambda_{1}\lesssim\textrm{max}\left(\lambda_{2}^{2},\,\lambda_{3}^{2}\right)$.

{The GUT field $\Sigma$ which is a $5\times5$ matrix
can diagonalized as }

{
\begin{equation}
\begin{aligned}\Sigma_{i}^{j} & =\delta_{i}^{j}\sigma_{i}\,,\quad i,\,j=1,...,5\,\\
\sum_{i=1}^{5}\sigma_{i} & =0\,.
\end{aligned}
\label{Sigma-diag}
\end{equation}
Various symmetry breaking patterns of $\text{SU}(5)$ were studied in
\cite{Magg:1979pf}, among which the one with $\textrm{SU}(5)\to\textrm{SU}(3)_{c}\times\textrm{SU}(2)_{L}\times\textrm{U}(1)_{Y}$
corresponds to }

{
\begin{equation}
\langle\Sigma\rangle=\sqrt{\frac{2}{15}}\sigma.\textrm{diag}\left(1,\,1,\,1,-\frac{3}{2},\,-\frac{3}{2}\right)\,,\label{GUTfieldVEV}
\end{equation}
where $\sigma$ is scalar field that emerges from spontaneous breaking
of $\text{SU}(5)$. Substituting it in (\ref{TreeSU5}) the equations
of motion for the $\sigma$ field read}

{
\[
\Box\sigma+\frac{\lambda_{c}}{4}\sigma^{3}-\frac{\lambda_{2}}{2}\sigma\phi^{2}=0\,,
\]
where $\lambda_{c}=a+\frac{7}{15}b$. Taking $\lambda_{2}\ll\lambda_{c}$, the $\sigma$ field quickly evolves to its local minimum 
of the potential given by
\begin{equation}
\sigma^{2}=\frac{2\lambda_{2}}{\lambda_{c}}\phi^{2}\,,\label{Vevsigma}
\end{equation} 
Adding the radiative corrections due to the couplings $-\frac{\lambda_{2}}{2}\phi^{2}\textrm{Tr}\Sigma^{2}$
and $\frac{\lambda_{3}}{2}\phi^{2}H_{5}^{\dagger}H_{5}$, 
the effective potential of $\phi$ gets to the CW form given by \cite{Linde:2005ht,Shafi:1983bd}

\begin{equation}
V_{eff}\left(\phi\right)=A\phi^{4}\left[\ln\left(\frac{\phi}{\mu}\right)+C\right]+V_{0}\,,\label{GUTCW}
\end{equation}
where 
\begin{equation}
A=\frac{\lambda_{2}^{2}}{16\pi^{2}}\left(1+\frac{25}{16}\frac{g^{4}}{\lambda_{c}^{2}}+\frac{14}{9}\frac{b^{2}}{\lambda_{c}}\right)\,.\label{AGUT}
\end{equation}
The $\left(\phi\,,\,\sigma\right)$ sector of effective potential
is given by

\begin{equation}
V_{eff}=\frac{\lambda_{c}}{16}\sigma^{4}-\frac{\lambda_{2}}{4}\sigma^{2}\phi^{2}+A\phi^{4}\left[\ln\left(\frac{\phi}{\mu}\right)+C\right]+V_{0}\,.\label{effphiad}
\end{equation}
and $\mu=\langle\phi\rangle$ denotes the VEV of $\phi$ at the minimum, $V_0$ and $C$ are dimensionfull and dimensionless constants respectively. Substituting (\ref{Vevsigma}) in (\ref{effphiad}) we obtain the effective potential for the field $\phi$ in the direction of $\sigma\propto \phi$. We set 
$V_{0}=\frac{A\mu^{4}}{4}$ which is the vacuum energy density i.e., $V\left(\phi=0\right)$ and the constant $C$ can be chosen such that $V\left(\phi=\mu\right)=0$.
Therefore, the potential (\ref{effphiad}) can be written as 

\begin{equation}
V_{eff}=A\phi^{4}\left[\ln\left(\frac{\phi}{\mu}\right)-\frac{1}{4}\right]+\frac{A\mu^{4}}{4}\,.\label{CWpot}
\end{equation}
{Following (\ref{Vevsigma}) the GUT field $\sigma$
reaches its global minimum only when the inflaton field reaches its VEV
by the end of inflation.} The inflationary predictions of
this model were reported in detail in \cite{Shafi:2006cs,Rehman:2008qs}. This
model was shown to be in good agreement with the spectral index $n_{s}=0.96-0.967$
and the tensor to scalar ratio $0.02\le r\le0.1$, which is well consistent
with the {\it Planck} 2015 data \cite{Ade:2015lrj,Okada:2014lxa}. 

From the VEV of the singlet field $\phi$ we can compute the masses
of superheavy gauge bosons as 

\begin{equation}
M_{X}=\sqrt{\frac{5\lambda_{2}g^{2}}{3\lambda_{c}A^{1/2}}}V_{0}^{1/4}\,.\label{Xboson}
\end{equation}
Taking $A\sim\frac{\lambda_{2}^{2}}{16\pi^{2}}$ the mass of gauge
bosons are approximately $2\sim4$ times larger than the scale
of vacuum energy $\left(V_{0}^{1/4}\right)$. The key prediction of
GUT models is proton decay $\left(p\to\pi^{0}+e^{+}\right)$ mediated
by $X,\,Y$ gauge bosons. The life time of proton can be computed
using 

\begin{equation}
\tau_{p}=\frac{M_{X}^{4}}{\alpha_{G}^{2}m_{pr}^{5}}\,,\label{prlifetime}
\end{equation}
where $m_{pr}$ is proton mass and $\alpha_{G}\sim1/40$ is the GUT
coupling constant. The current lower bound on proton life time is given
by $\tau_{p}>1.6\times10^{34}$ years indicates $M_{X}\sim4\times10^{15}\,\text{GeV}$
\cite{Nishino:2009aa,Miura:2016krn}. 

\section{GUT inflation with conformal symmetry }

\label{twofieldmodel-sec}

As discussed in Sec. \ref{SCmodel-sec}, conformal symmetry is useful
to generate flat potentials and the hierarchy of mass scales. Therefore,
embedding conformal symmetry in GUT inflation is more realistic and helpful to generate
simultaneously a Planck scale $m_{\rm P}$ along with the mass scale of X Bosons $M_{X}\sim10^{15}\,\text{GeV}$ that sources proton decay.
In this section, we extend the previously discussed CW inflation by
means of introducing conformal symmetry in SU(5) GUT theory. We then
obtain an interesting model of inflation by implementing spontaneous
breaking of conformal symmetry together with GUT symmetry\footnote{We note that conformal symmetry was considered in GUT inflation \cite{Esposito:1992xf,CervantesCota:1994zf,Buccella:1992rk}
but in those models the inflaton was a fundamental Higgs field of SU(5), whereas in our case the inflaton is a GUT singlet weakly coupled to the fundamental Higgs.}. We start with two complex singlet fields\footnote{A complex singlet is required to implement type I mechanism which we
later explain in Sec. \ref{seesawSec}. } of $\text{SU}(5)$ $\left(\Phi,\,\bar{X}\right)$ where the real
part of $\Phi$ ($\phi=\sqrt{2}\mathfrak{Re}\left[\Phi\right]$) is
identified as the inflaton. Gauge fixing the field $\bar{X}$ causes SBCS as discussed in Sec. \ref{SCmodel-sec}. It is worth to note here that the same framework we study here, based on $\text{SU}(5)$ GUT,
can be easily realized in the $\text{SO}(10)$ GUT. Therefore, the two complex singlets of $\text{SU}(5)$ considered here are also singlets of $\text{SO}(10)$ \cite{Lazarides:1991wu,Rehman:2008qs}. 

The conformally invariant action with complex SU(5) singlet fields
$\left(\Phi,\,\bar{X}\right)$ can be written as

\begin{equation}
\begin{split}S_{G}= & \int d^{4}x\,\sqrt{-g}\Bigg[\left(\vert\bar{X}\vert^{2}-\vert\Phi\vert^{2}-\textrm{Tr}\Sigma^{2}\right)\frac{R}{12}-\frac{1}{2}\left(\partial\Phi\right)^{\dagger}\left(\partial\Phi\right)+\frac{1}{2}\left(\partial\bar{X}\right)^{\dagger}\left(\partial\bar{X}\right)\\
 & -\frac{1}{2}\text{Tr}\left[\left(D^{\mu}\Sigma\right)^{\dagger}\left(D_{\mu}\Sigma\right)\right]-\frac{1}{4}\text{Tr}\left(\boldsymbol{F}_{\mu\nu}\boldsymbol{F}^{\mu\nu}\right)-V\left(\Phi,\,\bar{X},\,\Sigma\right)\Bigg]\,,
\end{split}
\label{CFTSU(5)}
\end{equation}
where $D_{\mu}\Sigma=\partial_{\mu}\Sigma-ig\left[\boldsymbol{A}_{\mu},\,\Sigma\right]$, $\boldsymbol{A}_{\mu}$ are the 24 massless Yang-Mills fields
with Field strength defined by $\boldsymbol{F}_{\mu\nu}\equiv\boldsymbol{\nabla}_{[\mu}\boldsymbol{A}_{\nu]}-ig\left[\boldsymbol{A}_{\mu},\,\boldsymbol{A}_{\nu}\right]$.
Here we assume the $\Phi$ field coupling to the Higgs field $H_{5}$ is negligible and not very relevant during
inflation. We consider that the singlet field $\Phi$ is weakly coupled
to the adjoint field $\Sigma$ through the following tree level potential

\begin{equation}
V\left(\Phi,\,\bar{X},\,\Sigma\right)=\frac{1}{4}a\left(\textrm{Tr}\Sigma^{2}\right)^{2}+\frac{1}{2}b\textrm{Tr}\Sigma^{4}-\frac{\lambda_{2}}{2}\vert\Phi\vert^{2}\textrm{Tr}\Sigma^{2}f\left(\frac{\Phi}{\bar{X}}\right)+\frac{\lambda_{1}}{4}\vert\Phi\vert^{4}f^{2}\left(\frac{\Phi}{\bar{X}}\right)\,,\label{potCFTSU5}
\end{equation}
where the coefficients $a\sim b\sim g^{2}$(gauge couplings $g^{2}\sim0.3$).
Following the discussion in section \ref{SCmodel-sec} we assume the
coupling constants are field dependent, i.e., in (\ref{potCFTSU5})
the coupling constants can be read as $\tilde{\lambda}_{2}=\lambda_{2}f\left(\frac{\Phi}{\bar{X}}\right),\,\tilde{\lambda}_{1}=\lambda_{1}f^{2}\left(\frac{\Phi}{\bar{X}}\right)$
which depend on the ratio of the fields $\left(\Phi,\,\bar{X}\right)$.
We consider 
\begin{equation}
f\left(\frac{\Phi}{\bar{X}}\right)=\left(1-\frac{\vert\Phi\vert^{2}}{\vert\bar{X}\vert^{2}}\right)\,.\label{fPchi}
\end{equation}
With the tree level potential in (\ref{potCFTSU5}), the action (\ref{CFTSU(5)})
is conformally invariant under the following transformations

\begin{equation}
g_{\mu\nu}\to\Omega\left(x\right)^{2}g_{\mu\nu}\quad,\quad\bar{X}\to\Omega^{-1}\left(x\right)\bar{X}\quad,\quad\Phi\to\Omega^{-1}\left(x\right)\Phi\quad,\quad\Sigma\to\Omega^{-1}\left(x\right)\Sigma\,.\label{ConformtrSU5}
\end{equation}
{The SBCS occurs
with gauge fixing $\bar{X}=\bar{X}^{*}=\sqrt{3}M$, where $M\sim\mathcal{O}\left(m_{\rm P}\right)$.
We assume inflation to happen in a direction $\text{I}m\Phi=0$.
Therefore, for the stability of inflaton trajectory we require the
mass of $\text{I}m\Phi$ to be}\footnote{{Where $H_{inf}$ is the Hubble parameter during inflation.}}{{}
$m_{\text{Im}\Phi}^{2}\gg H_{inf}^{2}$. To arrange this, we can add
a new term to the potential (\ref{potCFTSU5}) as }

{
\begin{equation}
V_{S}=V\left(\Phi,\,\bar{X},\,\Sigma\right)+\frac{\lambda_{im}}{4}\left(\Phi-\Phi^{\dagger}\right)^{2}\left(\Phi+\Phi^{\dagger}\right)^{2}\,,\label{stability-Pot}
\end{equation}
such that the mass of the $\text{Im}\Phi$ in the inflationary direction
is $m_{\text{Im}\Phi}^{2}=\frac{\partial^{2}V_{S}}{\partial\text{Im}\Phi^{2}}=\lambda_{im}\left(\Phi+\Phi^{*}\right)^{2}$.
Therefore, if $\lambda_{im}\gg\lambda_{1,2}$ we can have $m_{\text{Im}\Phi}^{2}\Big\vert_{\text{Im}\Phi=0}\gg H_{inf}^{2}$
during inflation. In this way, we can successfully obtain the stability
of the inflaton trajectory during inflation \cite{Kallosh:2010xz}. Note that when multiple non-minimally scalar fields are involved, it is in generally expected to induce geometrical destabilization effects due to the negative curvature of the fields space in the Einstein frame. This topic has been extensively studied in recent years \cite{Kaiser:2010ps,Renaux-Petel:2015mga,Brown:2017osf,Christodoulidis:2018qdw,Garcia-Saenz:2018ifx,Iarygina:2018kee}. In Appendix.~\ref{A1}, we present details of fields space geometry and argue that these effects might be negligible in the model we study herein, deferring a detailed quantitative analysis for future investigations.
 Similarly to the SV model, we also consider here $\textrm{SU}(5)\to\textrm{SU}(3)_{c}\times\textrm{SU}(2)_{L}\times\textrm{U}(1)_{Y}$
by}

{
\begin{equation}
\langle\Sigma\rangle=\sqrt{\frac{1}{15}}\sigma.\textrm{diag}\left(1,\,1,\,1,-\frac{3}{2},\,-\frac{3}{2}\right)\,.\label{GUTfieldVEV-1}
\end{equation}
Likewise to the SV model, we assume $\lambda_{1}\ll\lambda_{2}\ll a,\,b$
and due to the coupling $-\frac{\lambda_{2}}{2}\phi^{2}\textrm{Tr}\Sigma^{2}f\left(\frac{\phi}{\sqrt{6}M}\right)$,
the GUT field $\sigma$ reaches to its local field dependent minimum
given by}\footnote{{A similar scenario happens in the context of hybrid
inflationary scenario discussed in \cite{Buchmuller:2014dda}.}}

\begin{equation}
\sigma^{2}=\frac{2}{\lambda_{c}}\lambda_{2}\phi^{2}f\left(\frac{\phi}{\sqrt{6}M}\right)\,.\label{sigma2field}
\end{equation}
{Note that the above local minimum of the GUT field remains the same even though there is a non-minimal coupling with the Ricci scalar. We can easily understand this by conformally transforming the action (\ref{CFTSU(5)}) into the Einstein frame.}

After SU(5) symmetry breaking, the X gauge bosons become superheavy,
whereas the field $\sigma$ continues to follow the behavior of the
field $\phi$. The tree level potential for $\left(\phi,\,\sigma\right)$
sector is given by

\begin{equation}
V=\left[\frac{\lambda_{c}}{16}\sigma^{4}
-\frac{\lambda_{2}}{4}\sigma^{2}\phi^{2}f\left(\frac{\phi}{\sqrt{6}M}\right)
+\frac{\lambda_{1}}{4}\phi^{4}f^{2}\left(\frac{\phi}{\sqrt{6}M}\right)\right]\,.\label{treelevel3fieldpot}
\end{equation}
Substituting (\ref{sigma2field}) in (\ref{CFTSU(5)}) and
rescaling the field $\phi\to\sqrt{1+\frac{\lambda_{2}}{\lambda_{c}}}\phi$,
we obtain

\begin{equation}
\begin{aligned}S_{G}=\int d^{4}x\,\sqrt{-g}\Bigg\{ & \left(6M^{2}-\phi^{2}\right)\frac{R}{12}
-\frac{1}{2}\left(\partial\phi\right)^{2}\\
 & -\left[\frac{\lambda_{c}}{16}\sigma^{4}-\frac{\bar{\lambda}_{2}}{4}\sigma^{2}\phi^{2} f\left(\frac{\phi}{\sqrt{6}M}\right)+\frac{\bar{\lambda}_{1}}{4}\phi^{4}f^{2}\left(\frac{\phi}{\sqrt{6}M}\right)\right]\Bigg\}\,,
\end{aligned}
\label{actionphi}
\end{equation}
where $\bar{\lambda}_{1,2}=\lambda_{1,2}\sqrt{\frac{1}{1+\frac{\lambda_{2}}{\lambda_{c}}}}$.



Since $\lambda_{1}\ll\lambda_{2}$, the effective potential for the inflaton field $\phi$ due to the radiative corrections becomes 

\begin{equation}
V_{eff}\left(\phi\right)=V+\delta V+m_{\sigma}^{4}\ln\left(\frac{m_{\sigma}^{2}}{\mu^{2}}\right)+V_{0}\,,\label{vef}
\end{equation}
where $\delta V$ is the counter term, $\mu$ is the VEV of the field $\phi$ and $V_{0}$ is a constant.
Using (\ref{sigma2field}), choosing an appropriate $\delta V=\frac{\delta\bar{\lambda}_{2}}{4}\sigma^{2}\phi^{2}f^{2}\left(\frac{\phi}{\sqrt{6}M}\right)$,
a normalization constant such that $V_{eff}\left(\phi=\mu\right)=0$
and the vacuum energy density such that $V\left(\phi=0\right)=V_{0}=\frac{A\mu^{4}}{4}$,
we obtain

{
\begin{equation}
V_{eff}\left(\phi\right)=A\phi^{4}f^{2}\left(\frac{\phi}{\sqrt{6}M}\right)\ln\left(\left(\frac{6\phi^{2}M^{2}f\left(\frac{\phi}{\sqrt{6}M}\right)}
{\mu^{2}m_{\rm P}^{2}}\right)-\frac{1}{4}\right)+\frac{A\mu^{4}}{4}\,,\label{vef-1}
\end{equation}
where $A\sim\frac{\bar{\lambda}_{2}^2}{16\pi^{2}}$.}

We note here that the CW potential we considered is the standard one
obtained from 1-loop correction in Minkowski space-time. In the de
Sitter background, 1-loop corrections are in principle different and
their significance was discussed in literature \cite{Boyanovsky:2005px,Boyanovsky:2009xh,Destri:2009wn}.
Recently, in Ref.~\cite{Jain:2015jpa}, it was argued that during slow-roll
inflation we can neglect the contribution of 1-loop corrections in
the gravity sector. In addition, the contributions from higher loops
can also be neglected by the consideration of the slow-rolling scalar
field. Refs.~\cite{Kirsten:1993jn,Markkanen:2012rh}
provide quantum corrections calculated for the cases of non-minimally
coupled scalar fields.

{In order to get a Planck mass $m_{\rm P}$ dynamically generated by the end of inflation, we should take the corresponding
VEV of the inflaton field as}
\begin{equation}
\langle\phi\rangle=\mu=\sqrt{6}M\sqrt{1-\frac{m_{\rm P}^{2}}{M^2}}\,.\label{phiVev}
\end{equation}
We can see that $M\geq m_P$ and $\mu\to\sqrt{6}M$ if $M\gg m_P$. 

Taking the function $f\left(\frac{\phi}{\sqrt{6}M}\right)$ from (\ref{fPchi})
and by doing a conformal transformation of the action (\ref{actionphi}) with the effective potential (\ref{vef-1})
into Einstein frame, we obtain {(expressing in the
units of $m_{\rm P}=1$)}

{
\begin{equation}
S_{G}^{E}=\int d^{4}x\,\sqrt{-g_{E}}\Bigg[\frac{1}{2}R_{E}-\frac{1}{2M^{2}\left(1-\frac{\phi^{2}}{6M^{2}}\right)^{2}}\partial^{\mu}\phi\partial_{\mu}\phi-\frac{V_{eff}\left(\phi\right)}{36M^{4}f^{2}\left(\frac{\phi}{\sqrt{6}M}\right)}\Bigg]\,.\label{GUT starobinsky}
\end{equation}
Under the conformal transformation, the mass scales in the Einstein
frame must be redefined as $\mu^{2}\to\mu^{2}\left(6M^{2}-\phi^{2}\right)^{-1}$.
This is very much an equivalent procedure to the 1-loop analysis of Higgs
inflation. See Refs.~\cite{Bezrukov:2009db,George:2015nza,Fumagalli:2016lls,Pallis:2014cda}
for a detailed discussion on the equivalence between Jordan and Einstein
frames which exactly matches, if we redefine the mass scales accordingly
by conformal factor. Subsequently, substituting (\ref{vef-1}) in
(\ref{GUT starobinsky})}

{
\begin{equation}
S_{G}^{E}=\int d^{4}x\,\sqrt{-g}\left\{ \frac{1}{2}R_{E}-\frac{1}{2M^{2}\left(1-\frac{\phi^{2}}{6M^{2}}\right)^{2}}\partial^{\mu}\phi\partial_{\mu}\phi
-A\phi^{4}\left[\ln\left(\frac{\phi^{2}}{\mu^{2}}\right)-\frac{1}{4}\right]-\frac{A\mu^{4}}{4}\right\}\,.\label{conformalTaction}
\end{equation}
The kinetic term of (\ref{conformalTaction}) is similar the no-scale
models \cite{Ellis:2013nxa}.} Canonically normalizing the scalar
field as $\phi=\sqrt{6}M\tanh\left(\frac{\varphi}{\sqrt{6}}\right)$
yields the Einstein frame potential 

\begin{equation}
V_{E}\left(\varphi\right)=A\tanh^{4}\left(\frac{\varphi}{\sqrt{6}}\right)\left(\ln\left(\frac{6M^2\tanh^2\left(\frac{\varphi}{\sqrt{6}}\right)}{\mu^2}\right)-\frac{1}{4}\right)+\frac{A\mu^{4}}{4}\,.\label{varphipot}
\end{equation}
The corresponding VEV of the canonically normalized field is $\langle\varphi\rangle=\sqrt{6}\arctan\left(\frac{\mu}{\sqrt{6}M}\right)$.
The potential in (\ref{varphipot}) is a flattened version of CW potential
(\ref{CWpot}). Concretely, due to SBCS, the shape of the potential above VEV $\phi>\mu$
gets significantly flattened. In Fig. \ref{CWflat} we compare the
CW potential of the SV model with the modified form (\ref{varphipot})
we obtained in our case. The shape of the potential reaches a plateau
like in Starobinsky model when $\varphi\gg1$ i.e., $\phi\to\sqrt{6}M$.
{Inflation always starts near the plateau and continues
to evolve as $\phi\lesssim\sqrt{6}M$, therefore $f\left(\frac{\phi}{\sqrt{3}M}\right)$
defined in (\ref{fPchi}) is always positive and consequently that avoids an anti gravity regime.} {Note that
the flat potential (\ref{varphipot}) is significantly different
from the one of CW inflation, studied with positive non-minimal coupling
in \cite{Panotopoulos:2014hwa}.} {In the next
subsection we show that the inflationary observables for the potential
(\ref{varphipot}) exactly match those of Starobinsky and Higgs
inflation. }

\begin{figure}[h]
\centering\includegraphics[height=2.5in]{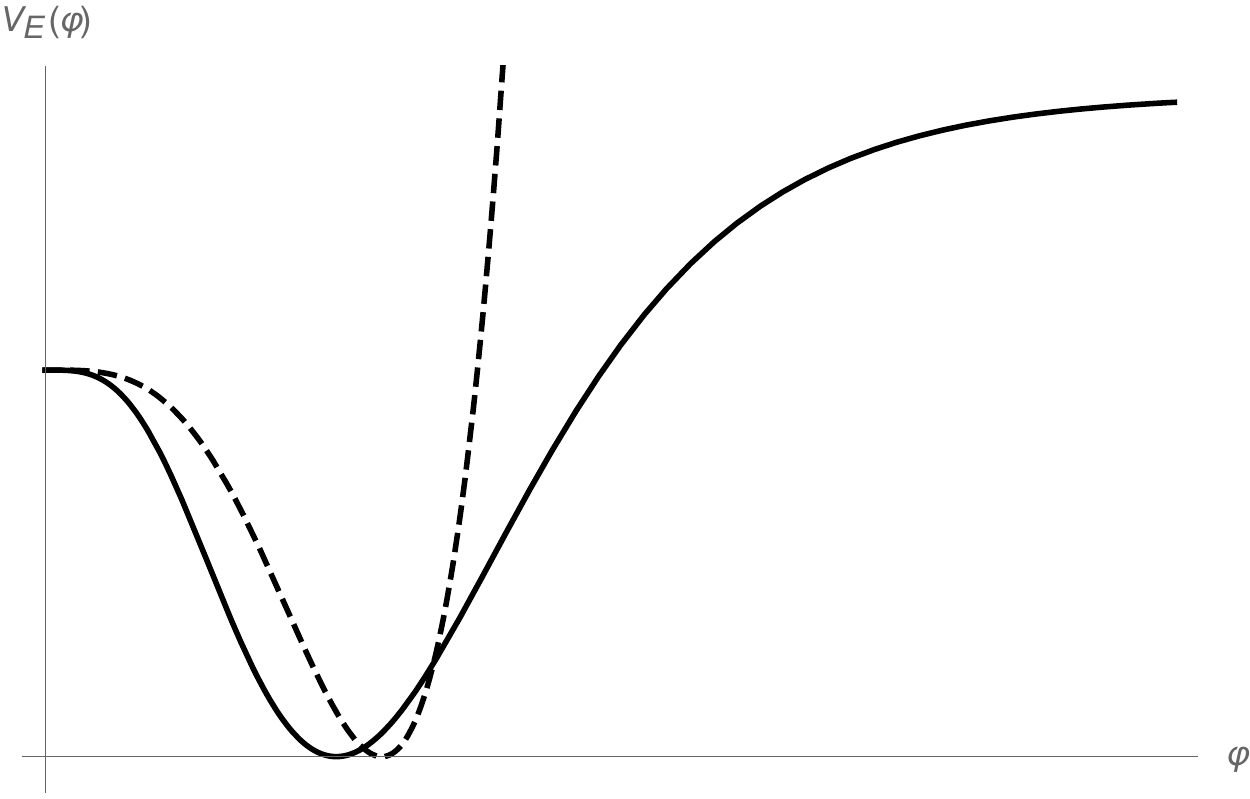}\caption{The dashed line denotes the CW potential in SV model. The full line
indicates the shape of the potential obtained in (\ref{varphipot}) which
comes from the insertion of conformal symmetry in SU(5). When $\varphi\gg\mu$
the above VEV (AV) branch of the potential approaches the plateau of Starobinsky
model. }

\label{CWflat} 
\end{figure}



\subsection{Inflationary predictions and proton lifetime}

We assume the standard Friedmann-Lema\^itre-Robertson-Walker (FLRW) background. Let us define the general definitions
of slow-roll parameters as 

\begin{equation}
\epsilon=\frac{H^{\prime}}{H}\quad,\quad\eta=-\frac{\epsilon^{\prime}}{\epsilon}\quad,\quad\zeta=-\frac{\eta^{\prime}}{\eta}\quad,\quad\delta=-\frac{\zeta^{\prime}}{\zeta}\,,\label{slwrolls}
\end{equation}
where $H$ is the Hubble parameter and the prime $^{\prime}$ denotes
derivative with respect to e-folding number $N=\ln a\left(t\right)$
before the end of inflation.
The scalar power spectrum is given by

\begin{equation}
{\cal P}_{\mathcal{R}}=\frac{\gamma_{s}H^{2}}{8\pi^{2}\epsilon}\Biggr|_{k=aH}\quad,\quad\gamma_{s}\equiv2^{2\nu_{s}-3}\frac{\Gamma\left(\nu_{s}\right){}^{2}}{\Gamma(3/2)^{2}}\left(1-\epsilon\right)^{2}\,.\label{pwrspectrum}
\end{equation}
The scalar power spectrum amplitude at pivot scale $k=0.002\,Mpc^{-1}$
is measured to be $P_{\mathcal{R}_{*}}=2.2\times10^{-9}$ \cite{Ade:2015lrj}.
The scalar spectral index up to the first orders in slow roll parameters
is given by 

\begin{equation}
n_{s}=1-2\epsilon-\eta\,.\label{ns}
\end{equation}
The running and running of running spectral index known as \cite{vandeBruck:2016rfv}

\begin{equation}
\begin{aligned}\alpha_{s}\equiv & \frac{dn_{s}}{d\ln k}\Biggr|_{k=aH}\simeq-2\epsilon\eta-\zeta\delta\,,\\
\beta_{s}\equiv & \frac{d\alpha_{s}}{d\ln k}\Biggr|_{k=aH}\simeq-2\epsilon\left(\eta+\zeta\right)-\eta\zeta\left(\zeta+\delta\right)\,.
\end{aligned}
\label{runnings}
\end{equation}
The ratio of tensor to scalar power spectrum is 

\begin{equation}
r=16\epsilon\Big\vert_{k=aH}\,.\label{r-}
\end{equation}
{The potential (\ref{varphipot}) when $\varphi\gg1$ (AV branch)
can be approximated as }

{
\begin{equation}
\begin{aligned}V_{E}\left(\varphi\right) & \simeq A\left(1-e^{-\sqrt{2/3}\varphi}\right)^{4}\ln\left(\frac{\sqrt{6}M\left(1-e^{-\sqrt{2/3}\varphi}\right)}{\mu}\right)\\
 & \approx A\left(1-e^{-\sqrt{2/3}\varphi}\right)^{4}\LT\ln\left(\frac{\sqrt{6}M}{\mu}\right)-e^{-\sqrt{2/3}\varphi}+\mathcal{O}\LF e^{-\sqrt{\frac{2}{3}}\varphi}\RF^2\RT\,.
\end{aligned}
\label{vaphiplotapp}
\end{equation}
The equation of motion of the canonically normalized field is }

{
\[
\ddot{\varphi}+3H\dot{\varphi}+V_{E,\varphi}=0\,,
\]
 which during the slow-roll regime reduces to }

{
\begin{equation}
\begin{aligned}
&\frac{\partial\varphi}{\partial N}\approx\frac{V_{E,\varphi}}{V_{E}}=\sqrt{\frac{2}{3}}\LF 4+\frac{1}{\ln\LF\frac{\sqrt{6}M}{\mu}\RF}\RF e^{-\sqrt{\frac{2}{3}}\varphi}\,\\
\implies& e^{\sqrt{\frac{2}{3}}\varphi(N_\ast)}= \frac{2}{3}\LF 4+\frac{1}{\ln\LF\frac{\sqrt{6}M}{\mu}\RF}\RF N_\ast
\label{slwapp}
\end{aligned}
\end{equation}
}
where we took $H_{inf}\approx\frac{V_{E}\left(\varphi\right)}{3}$ and $N_\ast$ is the $60$ $e-$ foldings before the end of inflation. 
Computing the slow-roll parameter using (\ref{slwapp}) we obtain 

\begin{equation}
\begin{aligned}
\epsilon&= \frac{\partial\ln H}{\partial N}\approx \frac{1}{2}\left(\frac{V_{E,\varphi}}{V_{E}}\right)^{2} \approx \frac{3}{4N^2}\,\\
\eta&=-\frac{\partial\epsilon}{\partial N}\approx\frac{2}{N}\,.
\end{aligned}
\label{epsilon}
\end{equation}

Using (\ref{epsilon}) we can write the predictions for the scalar
tilt (\ref{ns}) and tensor to scalar ratio (\ref{r-}) as }

{
\begin{equation}
n_{s}\approx1-\frac{2}{N}\,,\quad r=\frac{12}{N^{2}}\,,\label{ns-r-1}
\end{equation}
which exactly match with the predictions of Starobinsky and Higgs
inflation \cite{Starobinsky:1980te,Bezrukov:2007ep}. {We emphasize
that the predictions of our model in (\ref{ns-r-1}) are almost independent
of the VEV of the inflaton field $\langle\phi\rangle=\mu$.} }

In Table.~\ref{tab1} we present the inflationary predictions of the
model together with the corresponding $X$ bosons mass and proton
life time using (\ref{Xboson}) and (\ref{prlifetime}). We also show our results for the case when the inflaton field
rolls from above VEV (AV) i.e., when $\phi>\mu$. The predictions
of below VEV (BV) branch i.e., when $\phi<\mu$ are not very interesting
as those are nearly same in the original CW inflation without any
conformal symmetry \cite{Rehman:2008qs}. This is evident from Fig.~\ref{CWflat} where we can see only the AV branch of the potential
significantly different in our case, whereas the BV branch is nearly
same as in the SV model. Therefore, our interest in this paper
is restricted to AV branch. For this case, from Table. \ref{tab1}
we can see that the inflationary predictions of the model almost remains the same for any value of inflaton VEV. {Note that even though the inflaton field values are trans-Planckian, the values of $n_s$,\,$r$ remain the same. This is due to the fact that when $\varphi\gg \mu$ the shape of the potential is exponentially flat like in Starobinsky model. Therefore, inflationary predictions only depend on the potential plateau rather than the field values (shift symmetry).}
\selectlanguage{british}%
\begin{center}
\begin{table}[!h]
\centering{\scriptsize{}}%
\begin{tabular}{|c|c|c|c|c|c|c|c|c|c|c|c|c|c|c|c|}
\hline 
\selectlanguage{english}%
{\scriptsize{}\enskip{}$M$\enskip{}}\selectlanguage{british}%
 & \selectlanguage{english}%
{\scriptsize{}\enskip{}$\mu$\enskip{}}\selectlanguage{british}%
 & \selectlanguage{english}%
{\scriptsize{}\enskip{}$\langle\varphi\rangle$\enskip{}}\selectlanguage{british}%
 & \selectlanguage{english}%
{\scriptsize{}$\enskip A\enskip$}\selectlanguage{british}%
 & \selectlanguage{english}%
{\scriptsize{}$V_{0}^{1/4}$}\selectlanguage{british}%
 & \selectlanguage{english}%
{\scriptsize{}$V\left(\phi_{0}\right)^{1/4}$}\selectlanguage{british}%
 & \selectlanguage{english}%
{\scriptsize{}$H_{inf}$}\selectlanguage{british}%
 & \selectlanguage{english}%
{\scriptsize{}\enskip{}$N_{e}$\enskip{}}\selectlanguage{british}%
 & \selectlanguage{english}%
{\scriptsize{}\enskip{}$\varphi_{0}$\enskip{}}\selectlanguage{british}%
 & \selectlanguage{english}%
{\scriptsize{}\enskip{}$\varphi_{e}$\enskip{}}\selectlanguage{british}%
 & \selectlanguage{english}%
{\scriptsize{}$\enskip n_{s}\enskip$}\selectlanguage{british}%
 & \selectlanguage{english}%
{\scriptsize{}$\enskip r\enskip$}\selectlanguage{british}%
 & \selectlanguage{english}%
{\scriptsize{}\enskip{}$-\alpha_{s}$\enskip{}}\selectlanguage{british}%
 & \selectlanguage{english}%
{\scriptsize{}$\enskip-\beta_{s}\enskip$}\selectlanguage{british}%
 & \selectlanguage{english}%
{\scriptsize{}$\enskip M_{X}\enskip$}\selectlanguage{british}%
 & \selectlanguage{english}%
{\scriptsize{}$\enskip\tau_{p}\enskip$}\selectlanguage{british}%
\tabularnewline
\selectlanguage{english}%
{\scriptsize{}$\left(m_{\rm P}\right)$}\selectlanguage{british}%
 & \selectlanguage{english}%
{\scriptsize{}$\left(m_{\rm P}\right)$}\selectlanguage{british}%
 & \selectlanguage{english}%
{\scriptsize{}$\left(m_{\rm P}\right)$}\selectlanguage{british}%
 & \selectlanguage{english}%
{\scriptsize{}$\left(10^{-12}\right)$}\selectlanguage{british}%
 & \selectlanguage{english}%
{\scriptsize{}$\left(10^{16}\,\text{Gev}\right)$}\selectlanguage{british}%
 & \selectlanguage{english}%
{\scriptsize{}$\left(10^{16}\,\text{Gev}\right)$}\selectlanguage{british}%
 & \selectlanguage{english}%
{\scriptsize{}$\left(10^{13}\,\text{Gev}\right)$}\selectlanguage{british}%
 & \selectlanguage{english}%
\selectlanguage{british}%
 & \selectlanguage{english}%
{\scriptsize{}$\left(m_{\rm P}\right)$}\selectlanguage{british}%
 & \selectlanguage{english}%
{\scriptsize{}$\left(m_{\rm P}\right)$}\selectlanguage{british}%
 & \selectlanguage{english}%
\selectlanguage{british}%
 & \selectlanguage{english}%
\selectlanguage{british}%
 & \selectlanguage{english}%
{\scriptsize{}$\left(10^{-4}\right)$}\selectlanguage{british}%
 & \selectlanguage{english}%
{\scriptsize{}$\left(10^{-5}\right)$}\selectlanguage{british}%
 & \selectlanguage{english}%
{\scriptsize{}$\left(\sim10^{16}\,\text{Gev}\right)$}\selectlanguage{british}%
 & \selectlanguage{english}%
{\scriptsize{}$\left(\text{years}\right)$ }\selectlanguage{british}%
\tabularnewline
\hline 
\selectlanguage{english}%
{\scriptsize{}$1.1$}\selectlanguage{british}%
 & \selectlanguage{english}%
{\scriptsize{}$1.123$}\selectlanguage{british}%
 & \selectlanguage{english}%
{\scriptsize{}$1.09$}\selectlanguage{british}%
 & \selectlanguage{english}%
{\scriptsize{}$4.79$}\selectlanguage{british}%
 & \selectlanguage{english}%
{\scriptsize{}$0.29$}\selectlanguage{british}%
 & \selectlanguage{english}%
{\scriptsize{}$0.85$}\selectlanguage{british}%
 & \selectlanguage{english}%
{\scriptsize{}$1.74$}\selectlanguage{british}%
 & \selectlanguage{english}%
{\scriptsize{}$50$}\selectlanguage{british}%
 & \selectlanguage{english}%
{\scriptsize{}$7.24$}\selectlanguage{british}%
 & \selectlanguage{english}%
{\scriptsize{}$2.10$}\selectlanguage{british}%
 & \selectlanguage{english}%
{\scriptsize{}$0.960$}\selectlanguage{british}%
 & \selectlanguage{english}%
{\scriptsize{}$0.0048$}\selectlanguage{british}%
 & \selectlanguage{english}%
{\scriptsize{}$8.07$}\selectlanguage{british}%
 & \selectlanguage{english}%
{\scriptsize{}$2.67$}\selectlanguage{british}%
 & \selectlanguage{english}%
{\scriptsize{}$0.57$}\selectlanguage{british}%
 & \selectlanguage{english}%
{\scriptsize{}$5.0\times10^{34}$}\selectlanguage{british}%
\tabularnewline
\selectlanguage{english}%
\selectlanguage{british}%
 & \selectlanguage{english}%
\selectlanguage{british}%
 & \selectlanguage{english}%
\selectlanguage{british}%
 & \selectlanguage{english}%
{\scriptsize{}$3.95$}\selectlanguage{british}%
 & \selectlanguage{english}%
{\scriptsize{}$0.27$}\selectlanguage{british}%
 & \selectlanguage{english}%
{\scriptsize{}$0.82$}\selectlanguage{british}%
 & \selectlanguage{english}%
{\scriptsize{}$1.59$}\selectlanguage{british}%
 & \selectlanguage{english}%
{\scriptsize{}$55$}\selectlanguage{british}%
 & \selectlanguage{english}%
{\scriptsize{}$7.35$}\selectlanguage{british}%
 & \selectlanguage{english}%
{\scriptsize{}$2.10$}\selectlanguage{british}%
 & \selectlanguage{english}%
{\scriptsize{}$0.963$}\selectlanguage{british}%
 & \selectlanguage{english}%
{\scriptsize{}$0.0039$}\selectlanguage{british}%
 & \selectlanguage{english}%
{\scriptsize{}$6.67$}\selectlanguage{british}%
 & \selectlanguage{english}%
{\scriptsize{}$2.36$}\selectlanguage{british}%
 & \selectlanguage{english}%
{\scriptsize{}$0.54$}\selectlanguage{british}%
 & \selectlanguage{english}%
{\scriptsize{}$4.2\times10^{34}$}\selectlanguage{british}%
\tabularnewline
\selectlanguage{english}%
\selectlanguage{british}%
 & \selectlanguage{english}%
\selectlanguage{british}%
 & \selectlanguage{english}%
\selectlanguage{british}%
 & \selectlanguage{english}%
{\scriptsize{}$3.32$}\selectlanguage{british}%
 & \selectlanguage{english}%
{\scriptsize{}$0.26$}\selectlanguage{british}%
 & \selectlanguage{english}%
{\scriptsize{}$0.78$}\selectlanguage{british}%
 & \selectlanguage{english}%
{\scriptsize{}$1.46$}\selectlanguage{british}%
 & \selectlanguage{english}%
{\scriptsize{}$60$}\selectlanguage{british}%
 & \selectlanguage{english}%
{\scriptsize{}$7.46$}\selectlanguage{british}%
 & \selectlanguage{english}%
{\scriptsize{}$2.10$}\selectlanguage{british}%
 & \selectlanguage{english}%
{\scriptsize{}$0.966$}\selectlanguage{british}%
 & \selectlanguage{english}%
{\scriptsize{}$0.0033$}\selectlanguage{british}%
 & \selectlanguage{english}%
{\scriptsize{}$5.61$}\selectlanguage{british}%
 & \selectlanguage{english}%
{\scriptsize{}$2.90$}\selectlanguage{british}%
 & \selectlanguage{english}%
{\scriptsize{}$0.52$}\selectlanguage{british}%
 & \selectlanguage{english}%
{\scriptsize{}$3.6\times10^{34}$}\selectlanguage{british}%
\tabularnewline
\hline 
\selectlanguage{english}%
{\scriptsize{}$1.5$}\selectlanguage{british}%
 & \selectlanguage{english}%
{\scriptsize{}$2.738$}\selectlanguage{british}%
 & \selectlanguage{english}%
{\scriptsize{}$2.36$}\selectlanguage{british}%
 & \selectlanguage{english}%
{\scriptsize{}$6.87$}\selectlanguage{british}%
 & \selectlanguage{english}%
{\scriptsize{}$0.76$}\selectlanguage{british}%
 & \selectlanguage{english}%
{\scriptsize{}$0.85$}\selectlanguage{british}%
 & \selectlanguage{english}%
{\scriptsize{}$1.71$}\selectlanguage{british}%
 & \selectlanguage{english}%
{\scriptsize{}$50$}\selectlanguage{british}%
 & \selectlanguage{english}%
{\scriptsize{}$7.95$}\selectlanguage{british}%
 & \selectlanguage{english}%
{\scriptsize{}$3.093$}\selectlanguage{british}%
 & \selectlanguage{english}%
{\scriptsize{}$0.960$}\selectlanguage{british}%
 & \selectlanguage{english}%
{\scriptsize{}$0.0046$}\selectlanguage{british}%
 & \selectlanguage{english}%
{\scriptsize{}$7.88$}\selectlanguage{british}%
 & \selectlanguage{english}%
{\scriptsize{}$3.38$}\selectlanguage{british}%
 & \selectlanguage{english}%
{\scriptsize{}$1.53$}\selectlanguage{british}%
 & \selectlanguage{english}%
{\scriptsize{}$2.6\times10^{36}$}\selectlanguage{british}%
\tabularnewline
\selectlanguage{english}%
\selectlanguage{british}%
 & \selectlanguage{english}%
\selectlanguage{british}%
 & \selectlanguage{english}%
\selectlanguage{british}%
 & \selectlanguage{english}%
{\scriptsize{}$5.69$}\selectlanguage{british}%
 & \selectlanguage{english}%
{\scriptsize{}$0.72$}\selectlanguage{british}%
 & \selectlanguage{english}%
{\scriptsize{}$0.81$}\selectlanguage{british}%
 & \selectlanguage{english}%
{\scriptsize{}$1.56$}\selectlanguage{british}%
 & \selectlanguage{english}%
{\scriptsize{}$55$}\selectlanguage{british}%
 & \selectlanguage{english}%
{\scriptsize{}$8.07$}\selectlanguage{british}%
 & \selectlanguage{english}%
{\scriptsize{}$3.093$}\selectlanguage{british}%
 & \selectlanguage{english}%
{\scriptsize{}$0.964$}\selectlanguage{british}%
 & \selectlanguage{english}%
{\scriptsize{}$0.0038$}\selectlanguage{british}%
 & \selectlanguage{english}%
{\scriptsize{}$6.52$}\selectlanguage{british}%
 & \selectlanguage{english}%
{\scriptsize{}$2.27$}\selectlanguage{british}%
 & \selectlanguage{english}%
{\scriptsize{}$1.46$}\selectlanguage{british}%
 & \selectlanguage{english}%
{\scriptsize{}$2.1\times10^{36}$}\selectlanguage{british}%
\tabularnewline
\selectlanguage{english}%
\selectlanguage{british}%
 & \selectlanguage{english}%
\selectlanguage{british}%
 & \selectlanguage{english}%
\selectlanguage{british}%
 & \selectlanguage{english}%
{\scriptsize{}$4.79$}\selectlanguage{british}%
 & \selectlanguage{english}%
{\scriptsize{}$0.70$}\selectlanguage{british}%
 & \selectlanguage{english}%
{\scriptsize{}$0.77$}\selectlanguage{british}%
 & \selectlanguage{english}%
{\scriptsize{}$1.43$}\selectlanguage{british}%
 & \selectlanguage{english}%
{\scriptsize{}$60$}\selectlanguage{british}%
 & \selectlanguage{english}%
{\scriptsize{}$8.17$}\selectlanguage{british}%
 & \selectlanguage{english}%
{\scriptsize{}$3.093$}\selectlanguage{british}%
 & \selectlanguage{english}%
{\scriptsize{}$0.967$}\selectlanguage{british}%
 & \selectlanguage{english}%
{\scriptsize{}$0.0032$}\selectlanguage{british}%
 & \selectlanguage{english}%
{\scriptsize{}$5.48$}\selectlanguage{british}%
 & \selectlanguage{english}%
{\scriptsize{}$1.56$}\selectlanguage{british}%
 & \selectlanguage{english}%
{\scriptsize{}$1.39$}\selectlanguage{british}%
 & \selectlanguage{english}%
{\scriptsize{}$1.8\times10^{36}$}\selectlanguage{british}%
\tabularnewline
\hline 
\selectlanguage{english}%
{\scriptsize{}$2$}\selectlanguage{british}%
 & \selectlanguage{english}%
{\scriptsize{}$4.242$}\selectlanguage{british}%
 & \selectlanguage{english}%
{\scriptsize{}$3.23$}\selectlanguage{british}%
 & \selectlanguage{english}%
{\scriptsize{}$7.59$}\selectlanguage{british}%
 & \selectlanguage{english}%
{\scriptsize{}$1.21$}\selectlanguage{british}%
 & \selectlanguage{english}%
{\scriptsize{}$0.84$}\selectlanguage{british}%
 & \selectlanguage{english}%
{\scriptsize{}$1.70$}\selectlanguage{british}%
 & \selectlanguage{english}%
{\scriptsize{}$50$}\selectlanguage{british}%
 & \selectlanguage{english}%
{\scriptsize{}$8.63$}\selectlanguage{british}%
 & \selectlanguage{english}%
{\scriptsize{}$3.897$}\selectlanguage{british}%
 & \selectlanguage{english}%
{\scriptsize{}$0.960$}\selectlanguage{british}%
 & \selectlanguage{english}%
{\scriptsize{}$0.0045$}\selectlanguage{british}%
 & \selectlanguage{english}%
{\scriptsize{}$7.79$}\selectlanguage{british}%
 & \selectlanguage{english}%
{\scriptsize{}$3.47$}\selectlanguage{british}%
 & \selectlanguage{english}%
{\scriptsize{}$2.47$}\selectlanguage{british}%
 & \selectlanguage{english}%
{\scriptsize{}$1.6\times10^{37}$}\selectlanguage{british}%
\tabularnewline
\selectlanguage{english}%
\selectlanguage{british}%
 & \selectlanguage{english}%
\selectlanguage{british}%
 & \selectlanguage{english}%
\selectlanguage{british}%
 & \selectlanguage{english}%
{\scriptsize{}$6.29$}\selectlanguage{british}%
 & \selectlanguage{english}%
{\scriptsize{}$1.15$}\selectlanguage{british}%
 & \selectlanguage{english}%
{\scriptsize{}$0.80$}\selectlanguage{british}%
 & \selectlanguage{english}%
{\scriptsize{}$1.55$}\selectlanguage{british}%
 & \selectlanguage{english}%
{\scriptsize{}$55$}\selectlanguage{british}%
 & \selectlanguage{english}%
{\scriptsize{}$8.75$}\selectlanguage{british}%
 & \selectlanguage{english}%
{\scriptsize{}$3.897$}\selectlanguage{british}%
 & \selectlanguage{english}%
{\scriptsize{}$0.964$}\selectlanguage{british}%
 & \selectlanguage{english}%
{\scriptsize{}$0.0037$}\selectlanguage{british}%
 & \selectlanguage{english}%
{\scriptsize{}$6.45$}\selectlanguage{british}%
 & \selectlanguage{english}%
{\scriptsize{}$2.22$}\selectlanguage{british}%
 & \selectlanguage{english}%
{\scriptsize{}$2.30$}\selectlanguage{british}%
 & \selectlanguage{english}%
{\scriptsize{}$1.3\times10^{37}$}\selectlanguage{british}%
\tabularnewline
\selectlanguage{english}%
\selectlanguage{british}%
 & \selectlanguage{english}%
\selectlanguage{british}%
 & \selectlanguage{english}%
\selectlanguage{british}%
 & \selectlanguage{english}%
{\scriptsize{}$5.29$}\selectlanguage{british}%
 & \selectlanguage{english}%
{\scriptsize{}$1.11$}\selectlanguage{british}%
 & \selectlanguage{english}%
{\scriptsize{}$0.77$}\selectlanguage{british}%
 & \selectlanguage{english}%
{\scriptsize{}$1.52$}\selectlanguage{british}%
 & \selectlanguage{english}%
{\scriptsize{}$60$}\selectlanguage{british}%
 & \selectlanguage{english}%
{\scriptsize{}$8.85$}\selectlanguage{british}%
 & \selectlanguage{english}%
{\scriptsize{}$3.897$}\selectlanguage{british}%
 & \selectlanguage{english}%
{\scriptsize{}$0.967$}\selectlanguage{british}%
 & \selectlanguage{english}%
{\scriptsize{}$0.0032$}\selectlanguage{british}%
 & \selectlanguage{english}%
{\scriptsize{}$5.42$}\selectlanguage{british}%
 & \selectlanguage{english}%
{\scriptsize{}$1.48$}\selectlanguage{british}%
 & \selectlanguage{english}%
{\scriptsize{}$2.21$}\selectlanguage{british}%
 & \selectlanguage{english}%
{\scriptsize{}$1.1\times10^{37}$}\selectlanguage{british}%
\tabularnewline
\hline 
\selectlanguage{english}%
{\scriptsize{}$3$}\selectlanguage{british}%
 & \selectlanguage{english}%
{\scriptsize{}$6.928$}\selectlanguage{british}%
 & \selectlanguage{english}%
{\scriptsize{}$4.32$}\selectlanguage{british}%
 & \selectlanguage{english}%
{\scriptsize{}$7.92$}\selectlanguage{british}%
 & \selectlanguage{english}%
{\scriptsize{}$1.99$}\selectlanguage{british}%
 & \selectlanguage{english}%
{\scriptsize{}$0.84$}\selectlanguage{british}%
 & \selectlanguage{english}%
{\scriptsize{}$1.68$}\selectlanguage{british}%
 & \selectlanguage{english}%
{\scriptsize{}$50$}\selectlanguage{british}%
 & \selectlanguage{english}%
{\scriptsize{}$9.61$}\selectlanguage{british}%
 & \selectlanguage{english}%
{\scriptsize{}$5.956$}\selectlanguage{british}%
 & \selectlanguage{english}%
{\scriptsize{}$0.960$}\selectlanguage{british}%
 & \selectlanguage{english}%
{\scriptsize{}$0.0044$}\selectlanguage{british}%
 & \selectlanguage{english}%
{\scriptsize{}$7.73$}\selectlanguage{british}%
 & \selectlanguage{english}%
{\scriptsize{}$2.44$}\selectlanguage{british}%
 & \selectlanguage{english}%
{\scriptsize{}$3.99$}\selectlanguage{british}%
 & \selectlanguage{english}%
{\scriptsize{}$1.2\times10^{38}$}\selectlanguage{british}%
\tabularnewline
\selectlanguage{english}%
{\scriptsize{}$ $}\selectlanguage{british}%
 & \selectlanguage{english}%
{\scriptsize{}$ $}\selectlanguage{british}%
 & \selectlanguage{english}%
{\scriptsize{}$ $}\selectlanguage{british}%
 & \selectlanguage{english}%
{\scriptsize{}$6.57$}\selectlanguage{british}%
 & \selectlanguage{english}%
{\scriptsize{}$1.90$}\selectlanguage{british}%
 & \selectlanguage{english}%
{\scriptsize{}$0.80$}\selectlanguage{british}%
 & \selectlanguage{english}%
{\scriptsize{}$1.53$}\selectlanguage{british}%
 & \selectlanguage{english}%
{\scriptsize{}$55$}\selectlanguage{british}%
 & \selectlanguage{english}%
{\scriptsize{}$9.72$}\selectlanguage{british}%
 & \selectlanguage{english}%
{\scriptsize{}$5.956$}\selectlanguage{british}%
 & \selectlanguage{english}%
{\scriptsize{}$0.964$}\selectlanguage{british}%
 & \selectlanguage{english}%
{\scriptsize{}$0.0037$}\selectlanguage{british}%
 & \selectlanguage{english}%
{\scriptsize{}$6.40$}\selectlanguage{british}%
 & \selectlanguage{english}%
{\scriptsize{}$1.99$}\selectlanguage{british}%
 & \selectlanguage{english}%
{\scriptsize{}$3.81$}\selectlanguage{british}%
 & \selectlanguage{english}%
{\scriptsize{}$1\times10^{38}$}\selectlanguage{british}%
\tabularnewline
\selectlanguage{english}%
{\scriptsize{}$ $}\selectlanguage{british}%
 & \selectlanguage{english}%
{\scriptsize{}$ $}\selectlanguage{british}%
 & \selectlanguage{english}%
{\scriptsize{}$ $}\selectlanguage{british}%
 & \selectlanguage{english}%
{\scriptsize{}$5.54$}\selectlanguage{british}%
 & \selectlanguage{english}%
{\scriptsize{}$1.82$}\selectlanguage{british}%
 & \selectlanguage{english}%
{\scriptsize{}$0.77$}\selectlanguage{british}%
 & \selectlanguage{english}%
{\scriptsize{}$1.41$}\selectlanguage{british}%
 & \selectlanguage{english}%
{\scriptsize{}$60$}\selectlanguage{british}%
 & \selectlanguage{english}%
{\scriptsize{}$9.82$}\selectlanguage{british}%
 & \selectlanguage{english}%
{\scriptsize{}$5.956$}\selectlanguage{british}%
 & \selectlanguage{english}%
{\scriptsize{}$0.967$}\selectlanguage{british}%
 & \selectlanguage{english}%
{\scriptsize{}$0.0031$}\selectlanguage{british}%
 & \selectlanguage{english}%
{\scriptsize{}$5.39$}\selectlanguage{british}%
 & \selectlanguage{english}%
{\scriptsize{}$1.24$}\selectlanguage{british}%
 & \selectlanguage{english}%
{\scriptsize{}$3.65$}\selectlanguage{british}%
 & \selectlanguage{english}%
{\scriptsize{}$8.5\times10^{37}$}\selectlanguage{british}%
\tabularnewline
\hline 
\selectlanguage{english}%
{\scriptsize{}$5$}\selectlanguage{british}%
 & \selectlanguage{english}%
{\scriptsize{}$12$}\selectlanguage{british}%
 & \selectlanguage{english}%
{\scriptsize{}$5.62$}\selectlanguage{british}%
 & \selectlanguage{english}%
{\scriptsize{}$8.07$}\selectlanguage{british}%
 & \selectlanguage{english}%
{\scriptsize{}$3.47$}\selectlanguage{british}%
 & \selectlanguage{english}%
{\scriptsize{}$0.84$}\selectlanguage{british}%
 & \selectlanguage{english}%
{\scriptsize{}$1.68$}\selectlanguage{british}%
 & \selectlanguage{english}%
{\scriptsize{}$50$}\selectlanguage{british}%
 & \selectlanguage{english}%
{\scriptsize{}$12.5$}\selectlanguage{british}%
 & \selectlanguage{english}%
{\scriptsize{}$7.95$}\selectlanguage{british}%
 & \selectlanguage{english}%
{\scriptsize{}$0.960$}\selectlanguage{british}%
 & \selectlanguage{english}%
{\scriptsize{}$0.0044$}\selectlanguage{british}%
 & \selectlanguage{english}%
{\scriptsize{}$7.69$}\selectlanguage{british}%
 & \selectlanguage{english}%
{\scriptsize{}$3.04$}\selectlanguage{british}%
 & \selectlanguage{english}%
{\scriptsize{}$6.95$}\selectlanguage{british}%
 & \selectlanguage{english}%
{\scriptsize{}$7.8\times10^{38}$}\selectlanguage{british}%
\tabularnewline
\selectlanguage{english}%
{\scriptsize{}$ $}\selectlanguage{british}%
 & \selectlanguage{english}%
{\scriptsize{}$ $}\selectlanguage{british}%
 & \selectlanguage{english}%
{\scriptsize{}$ $}\selectlanguage{british}%
 & \selectlanguage{english}%
{\scriptsize{}$6.70$}\selectlanguage{british}%
 & \selectlanguage{english}%
{\scriptsize{}$3.32$}\selectlanguage{british}%
 & \selectlanguage{english}%
{\scriptsize{}$0.80$}\selectlanguage{british}%
 & \selectlanguage{english}%
{\scriptsize{}$1.53$}\selectlanguage{british}%
 & \selectlanguage{english}%
{\scriptsize{}$55$}\selectlanguage{british}%
 & \selectlanguage{english}%
{\scriptsize{}$12.7$}\selectlanguage{british}%
 & \selectlanguage{english}%
{\scriptsize{}$7.95$}\selectlanguage{british}%
 & \selectlanguage{english}%
{\scriptsize{}$0.964$}\selectlanguage{british}%
 & \selectlanguage{english}%
{\scriptsize{}$0.0037$}\selectlanguage{british}%
 & \selectlanguage{english}%
{\scriptsize{}$6.37$}\selectlanguage{british}%
 & \selectlanguage{english}%
{\scriptsize{}$2.51$}\selectlanguage{british}%
 & \selectlanguage{english}%
{\scriptsize{}$6.63$}\selectlanguage{british}%
 & \selectlanguage{english}%
{\scriptsize{}$9.2\times10^{38}$}\selectlanguage{british}%
\tabularnewline
\selectlanguage{english}%
{\scriptsize{}$ $}\selectlanguage{british}%
 & \selectlanguage{english}%
{\scriptsize{}$ $}\selectlanguage{british}%
 & \selectlanguage{english}%
{\scriptsize{}$ $}\selectlanguage{british}%
 & \selectlanguage{english}%
{\scriptsize{}$5.65$}\selectlanguage{british}%
 & \selectlanguage{english}%
{\scriptsize{}$3.18$}\selectlanguage{british}%
 & \selectlanguage{english}%
{\scriptsize{}$0.77$}\selectlanguage{british}%
 & \selectlanguage{english}%
{\scriptsize{}$1.41$}\selectlanguage{british}%
 & \selectlanguage{english}%
{\scriptsize{}$60$}\selectlanguage{british}%
 & \selectlanguage{english}%
{\scriptsize{}$12.8$}\selectlanguage{british}%
 & \selectlanguage{english}%
{\scriptsize{}$7.95$}\selectlanguage{british}%
 & \selectlanguage{english}%
{\scriptsize{}$0.967$}\selectlanguage{british}%
 & \selectlanguage{english}%
{\scriptsize{}$0.0031$}\selectlanguage{british}%
 & \selectlanguage{english}%
{\scriptsize{}$5.35$}\selectlanguage{british}%
 & \selectlanguage{english}%
{\scriptsize{}$1.97$}\selectlanguage{british}%
 & \selectlanguage{english}%
{\scriptsize{}$6.35$}\selectlanguage{british}%
 & \selectlanguage{english}%
{\scriptsize{}$1.3\times10^{40}$}\selectlanguage{british}%
\tabularnewline
\hline 
\selectlanguage{english}%
{\scriptsize{}$10$}\selectlanguage{british}%
 & \selectlanguage{english}%
{\scriptsize{}$24.37$}\selectlanguage{british}%
 & \selectlanguage{english}%
{\scriptsize{}$7.33$}\selectlanguage{british}%
 & \selectlanguage{english}%
{\scriptsize{}$8.13$}\selectlanguage{british}%
 & \selectlanguage{english}%
{\scriptsize{}$7.07$}\selectlanguage{british}%
 & \selectlanguage{english}%
{\scriptsize{}$0.84$}\selectlanguage{british}%
 & \selectlanguage{english}%
{\scriptsize{}$1.68$}\selectlanguage{british}%
 & \selectlanguage{english}%
{\scriptsize{}$50$}\selectlanguage{british}%
 & \selectlanguage{english}%
{\scriptsize{}$12.5$}\selectlanguage{british}%
 & \selectlanguage{english}%
{\scriptsize{}$7.95$}\selectlanguage{british}%
 & \selectlanguage{english}%
{\scriptsize{}$0.960$}\selectlanguage{british}%
 & \selectlanguage{english}%
{\scriptsize{}$0.0044$}\selectlanguage{british}%
 & \selectlanguage{english}%
{\scriptsize{}$7.68$}\selectlanguage{british}%
 & \selectlanguage{english}%
{\scriptsize{}$5.20$}\selectlanguage{british}%
 & \selectlanguage{english}%
{\scriptsize{}$14.1$}\selectlanguage{british}%
 & \selectlanguage{english}%
{\scriptsize{}$1.9\times10^{40}$}\selectlanguage{british}%
\tabularnewline
\hline 
\selectlanguage{english}%
{\scriptsize{}$ $}\selectlanguage{british}%
 & \selectlanguage{english}%
{\scriptsize{}$ $}\selectlanguage{british}%
 & \selectlanguage{english}%
{\scriptsize{}$ $}\selectlanguage{british}%
 & \selectlanguage{english}%
{\scriptsize{}$6.75$}\selectlanguage{british}%
 & \selectlanguage{english}%
{\scriptsize{}$6.75$}\selectlanguage{british}%
 & \selectlanguage{english}%
{\scriptsize{}$0.80$}\selectlanguage{british}%
 & \selectlanguage{english}%
{\scriptsize{}$1.53$}\selectlanguage{british}%
 & \selectlanguage{english}%
{\scriptsize{}$55$}\selectlanguage{british}%
 & \selectlanguage{english}%
{\scriptsize{}$12.7$}\selectlanguage{british}%
 & \selectlanguage{english}%
{\scriptsize{}$7.95$}\selectlanguage{british}%
 & \selectlanguage{english}%
{\scriptsize{}$0.964$}\selectlanguage{british}%
 & \selectlanguage{english}%
{\scriptsize{}$0.0037$}\selectlanguage{british}%
 & \selectlanguage{english}%
{\scriptsize{}$6.35$}\selectlanguage{british}%
 & \selectlanguage{english}%
{\scriptsize{}$4.04$}\selectlanguage{british}%
 & \selectlanguage{english}%
{\scriptsize{}$13.5$}\selectlanguage{british}%
 & \selectlanguage{english}%
{\scriptsize{}$1.6\times10^{40}$}\selectlanguage{british}%
\tabularnewline
\hline 
\selectlanguage{english}%
{\scriptsize{}$ $}\selectlanguage{british}%
 & \selectlanguage{english}%
{\scriptsize{}$ $}\selectlanguage{british}%
 & \selectlanguage{english}%
{\scriptsize{}$ $}\selectlanguage{british}%
 & \selectlanguage{english}%
{\scriptsize{}$5.69$}\selectlanguage{british}%
 & \selectlanguage{english}%
{\scriptsize{}$6.47$}\selectlanguage{british}%
 & \selectlanguage{english}%
{\scriptsize{}$0.77$}\selectlanguage{british}%
 & \selectlanguage{english}%
{\scriptsize{}$1.41$}\selectlanguage{british}%
 & \selectlanguage{english}%
{\scriptsize{}$60$}\selectlanguage{british}%
 & \selectlanguage{english}%
{\scriptsize{}$12.8$}\selectlanguage{british}%
 & \selectlanguage{english}%
{\scriptsize{}$7.95$}\selectlanguage{british}%
 & \selectlanguage{english}%
{\scriptsize{}$0.967$}\selectlanguage{british}%
 & \selectlanguage{english}%
{\scriptsize{}$0.0031$}\selectlanguage{british}%
 & \selectlanguage{english}%
{\scriptsize{}$5.33$}\selectlanguage{british}%
 & \selectlanguage{english}%
{\scriptsize{}$3.96$}\selectlanguage{british}%
 & \selectlanguage{english}%
{\scriptsize{}$12.9$}\selectlanguage{british}%
 & \selectlanguage{english}%
{\scriptsize{}$1.3\times10^{40}$}\selectlanguage{british}%
\tabularnewline
\hline 
\end{tabular}\caption{\foreignlanguage{english}{Inflationary predictions of the AV branch solutions for different
parameter values. }}

\label{tab1} 
\end{table}
\par\end{center}

\begin{figure}[H]
\centering\includegraphics[height=2.5in]{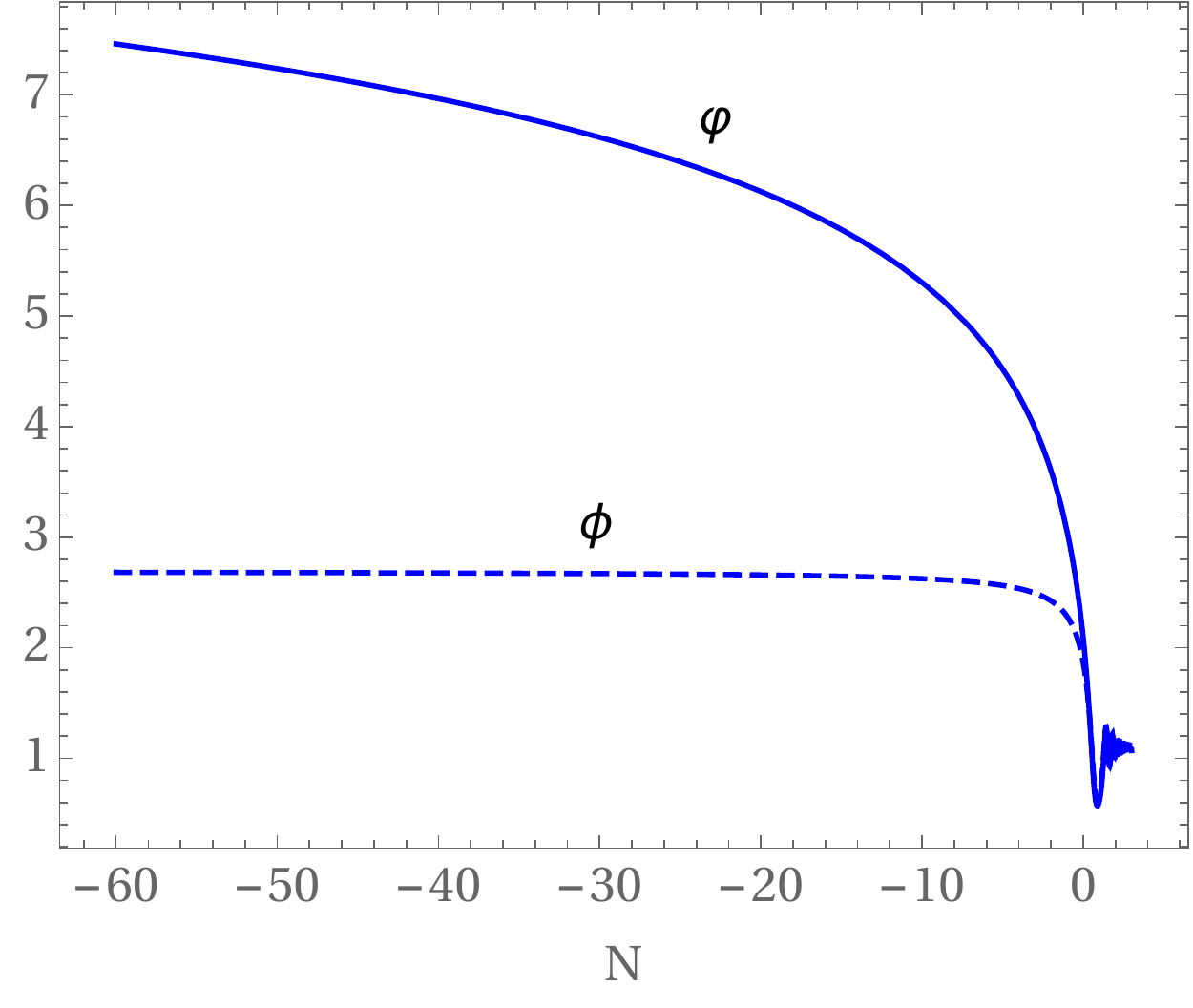}\quad{}\includegraphics[height=2.5in]{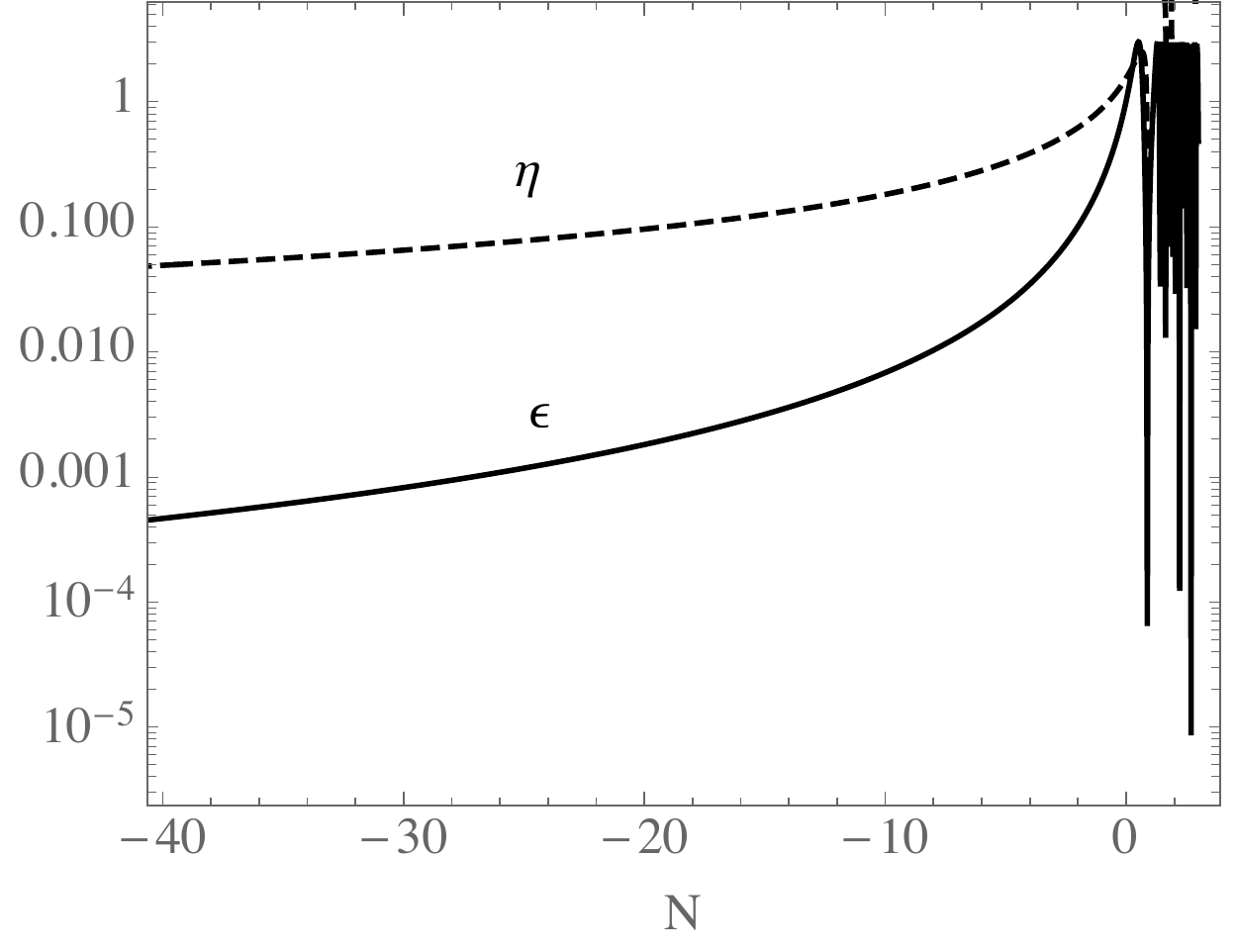}\caption{In the left panel we depict the evolution of the scalar field ({in the units of $m_P$}) during inflation
versus the e-folding number. The solid blue line indicates the evolution
of the canonically normalized field $\varphi$, whereas the dotted blue
line is for the original field $\phi$. In the right panel we plot
the corresponding slow-roll parameters $\epsilon,\,\eta$ versus $N$. Inflation
ends when $\epsilon,\,\eta=1$. For both plots, we have taken $\mu=1.12\,m_{\rm P}$. }

\label{fieldevolution} 
\end{figure}

\selectlanguage{english}%
In Fig.~\ref{fieldevolution} we depict the evolution of field $\phi$
(also for the canonically normalized field $\varphi$) and slow-roll parameter
$\epsilon$ for particular parameter values. 

\section{Type I seesaw mechanism and neutrino masses}

\label{seesawSec}

In this section, we further extend our model through a type I seesaw mechanism with global lepton number symmetry, whose spontaneous breaking leads to the generation of neutrino masses. In this framework, we suppose the singlet field $\Phi$ carries two units of lepton number
and is coupled to the three generation of singlet right handed Majorana neutrinos (RHNs), from \cite{Boucenna:2014uma}

\begin{equation}
V_{N}=V\left(\Phi,\,\bar{X},\,\Sigma\right)+Y_{D}^{ij}\bar{l_{L}^{j}}i\tau_{2}H^{\star}\nu_{R}^{i}+\frac{1}{2}Y_{N}^{i}\Phi f\left(\frac{\Phi}{\bar{X}}\right)  \overline{\nu_{R}^{ic}}\nu_{R}^{i}+h.c,\label{neutrino-1}
\end{equation}
where $l$ is the lepton doublet, $\tau_{2}$ is the second Pauli
matrix. Here $Y_{D}$ is the Yukawa coupling matrix of the SM Higgs
coupling to the left handed neutrinos and $Y_{N}$ is the coupling
matrix of the singlet field to the three generations of Majorana RHNs $\left(\nu_{R}^{i}\right)$. In principle, we can
also weakly couple the inflaton with the SM Higgs boson as

\begin{equation}
V_{h}=V_{N}+\lambda_{h}f\left(\frac{\Phi}{\bar{X}}\right)\Phi^{\dagger}\Phi H^{\dagger}H\,.\label{Reheapot}
\end{equation}
We note that even with the new potential in (\ref{Reheapot}), conformal
symmetry in (\ref{CFTSU(5)}) can be preserved by the following additional transformations\footnote{The kinetic terms and couplings of SM Higgs and RHNs to the Ricci scalar are irrelevant here and can be neglected in comparison with the inflaton dynamics.}

\begin{equation}
l_{L}^{i}\to\Omega^{3/2}l_{L}^{i}\,,\quad\nu_{R}^{i}\to\Omega^{3/2}\nu_{R}^{i}\,,\quad H\to\Omega H\label{newcon}
\end{equation}
Applying SBCS via $\bar{X}=\bar{X}^*=\sqrt{3}M$ and computing 1-loop corrections due to the additional couplings to neutrinos
(\ref{neutrino-1}) and SM Higgs, the effective potential of the field $\phi$ becomes

\begin{equation}
V_{f}^{eff}=\frac{36A_{f}M^{4}}{m_{\rm P}^{4}}f^{2}\left(\frac{\phi}{\sqrt{3}M}\right)\phi^{4}\ln\left(\frac{\phi^{2}f\left(\frac{\phi}{\sqrt{3}M}\right)}{\mu_{f}^{2}}
-\frac{1}{4}\right)+\frac{A_{f}\mu_{f}^{4}}{4}\,,\label{v1eff}
\end{equation}
where $A_{f}=\frac{\beta_{f}}{32\pi^{2}}$ and
\begin{equation}
\beta_{f}=20\bar{\lambda}_{2}^{2}+2\lambda_{h}^{2}+2\bar{\lambda}_{2}\sum_{i}\left(Y_{N}^{i}\right)^{2}-\sum_{i}\left(Y_{N}^{i}\right)^{4}\,.\label{betaf}
\end{equation}
In (\ref{betaf}) we assume the coupling constant $Y_{N}^{i}$ to
be at least $\mathcal{O}\left(10\right)$ smaller than $\bar{\lambda}_{2}$
and $\lambda_{h}\ll Y_{N}^{i}$, such that $\beta_{f}\sim20\bar{\lambda}_{2}^{2}$
and $\mu_{f}\sim\mu$. Therefore, during inflation the coupling of a
singlet field to the adjoint scalar $\Sigma$ dominates. Consequently,
the inflationary predictions in Table.~\ref{tab1} are unaffected by
these additional couplings to Higgs and singlet neutrinos. Moreover,  since we impose $\lambda_{h}\ll Y_{N}^{i}$, the inflaton field dominantly decays to RHNs rather than to SM Higgs. 

Lets consider that the lepton number violation happens at a scale
when $\langle\phi\rangle=\mu$. Computing the mass matrix of singlet
and doublet neutrinos in the basis of $\nu_{L},\,\nu_{R}$, using the
Einstein frame potential of (\ref{neutrino-1}), we have

\begin{equation}
\mathcal{M}_{\nu}=\left[\begin{matrix}0 & Y_{D}v_{2}\\
Y_{D}^{T}v_{2} & \frac{m_{\rm P}^{2}}{M^{2}}\frac{\langle\phi\rangle Y_{N}}{\sqrt{2}}
\end{matrix}\right]\,,\label{massmatrix}
\end{equation}
where $v_{2}=246\,\text{GeV}$ is the electroweak vacuum. The light left handed neutrino
mass can be obtained from perturbative diagonalization of (\ref{massmatrix})
as

\begin{equation}
m_{\nu_{L}}\simeq\sqrt{2}Y_{D}Y_{N}^{-1}Y_{D}^{T}\frac{v_{2}^{2}}{\mu}\frac{M^{2}}{m_{\rm P}^{2}}\,.\label{lightneutrino}
\end{equation}
The mass of heavy RHNs is given by 
\begin{equation}
m_{\nu_{R}}=\frac{Y_{N}\langle\phi\rangle}{\sqrt{2}}\frac{m_{\rm P}^{2}}{M^{2}}\,.\label{RHmass}
\end{equation}
The essence of the seesaw mechanism is the generation of neutrino masses, 
resulting in light left handed neutrinos and heavy right handed neutrinos.
Both are related here  to the VEV of the inflaton field. 

The current Planck data indicates the sum of light neutrino masses
constrained as $\sum m_{\nu_{i}}<0.23\,eV$ \cite{Ade:2015xua}. Therefore
considering the light neutrino mass to be $m_{\nu_{L}}\sim\mathcal{O}(0.1)\,\text{eV}$,
(\ref{lightneutrino}) gives a relation

\begin{equation}
Y_{N}\simeq6\sqrt{2}Y_{D}^{2}\frac{10^{14}\,GeV}{\mu}\frac{M^{2}}{m_{\rm P}^{2}}\,,\label{YNYD}
\end{equation}
Taking $Y_{D}\sim\mathcal{O}\left(10^{-1}\right)$ and from Table.~\ref{tab1} imposing $\mu\sim1.2\,m_{\rm P}-24.37\,m_{\rm P}$, we get $2.5\times10^{-6}\lesssim Y_{N}^{i}\lesssim1.0\times10^{-5}$.
This supports our previous assumptions after (\ref{betaf}), that the
couplings to the RHNs have a negligible effect for inflation. Our generalization of the SV model successfully fits into explaining the origin of neutrino masses. We can also take $Y_{D}<\mathcal{O}\left(10^{-1}\right)$
which results in smaller values for $Y_{N}<\mathcal{O}\left(10^{-6}\right)$.
Taking $Y_{N}\sim10^{-6}$, the heavy RHN mass will
be around $m_{\nu_{R}}\sim4\times10^{12}\,\text{GeV}.$ For $Y_{N}<\mathcal{O}\left(10^{-6}\right)$
we can lower the masses of RHNs. In the next section
we aim to study reheating in our inflationary scenario, taking into
account the constraints we have derived so far. 

\section{Reheating and non-thermal leptogenesis}

\label{ReheatSec}

We consider reheating through a dominant decay of the inflaton into heavy
RHNs\footnote{The inflaton could also decay into Higgs field but we have chosen the coupling of the Higgs field to the inflaton as $\lambda_h\ll Y_N^i \lesssim \mathcal{O}\LF 10^{-6} \RF$. For these couplings, the decay rate of the inflaton to a pair of Higgs bosons is negligible \cite{Kofman:1997yn,Greene:1997fu}. However, there can be a period of parametric resonance in the phase of preheating right after the end of inflation, during which the number of Higgs particles can grow exponentially \cite{Kofman:1997yn,Greene:1997fu}. Around the VEV, the inflaton potential (\ref{varphipot}) can be approximated as
	\begin{equation*}
	V_E\LF \varphi \RF = \frac{1}{2}m_\varphi^2\LF \varphi-\langle \varphi \rangle \RF^2 = \frac{1}{2}m^2\hat{\varphi}^2  \,. 
	\end{equation*}
Then we can apply the results of \cite{Kofman:1997yn,Greene:1997fu} to estimate the effect of parametric resonance. The inflaton field oscillates around the minimum as 
\begin{equation*}
\hat{\varphi}(t) \approx  \hat{\varphi}_A(t)\sin(mt)\,, 
\end{equation*}
where $\hat{\varphi}_A\LF t \RF \approx \frac{m_p}{\sqrt{3\pi}mt}$ is the amplitude of oscillations of the inflaton field. The regime of parametric resonance occurs as far as $\hat{\varphi}_A > \frac{\lambda_h^2}{8\pi}\langle \varphi \rangle$ and when $\hat{\varphi}_A$ drops to smaller values then standard perturbation theory dominates. To estimate the effect of parametric resonance in our case we compute the number of oscillations at the end of parametric resonance ($N_f$). {Following estimates from \cite{Kofman:1997yn} we especially have $N_f\approx \frac{mt_f}{2\pi}$ where $t_f$ is the instant when parametric resonance ends, by means of
\begin{equation*}
\lambda_h\hat{\varphi}_A\approx \frac{\lambda_hM_p}{3m_{\varphi}t_f} \approx m \,. 
\end{equation*}
As a result, we can further obtain \cite{Kofman:1997yn}
\begin{equation*}
N_f \sim \frac{\lambda_hm_P}{6\pi m_{\varphi}} \ll 1\,,
\end{equation*}
 since $m_\varphi \sim \mathcal{O}\LF 10^{-6} \RF$ from (\ref{infmas}) and $\lambda_h\ll 10^{-6}$ in our case.} Therefore, the effects of parametric resonance in our case is negligible for our chosen values of inflaton-Higgs couplings.}\footnote{We ignore the effects of non-minimally coupled heavy fields $\text{Im}[\Phi],\,\sigma$ during preheating or reheating due to non-trivial fields space geometry in the Einstein frame \cite{Ema:2016dny,DeCross:2016fdz,Krajewski:2018moi,Iarygina:2018kee}. We defer these interesting studies for future investigation. }which requires $m_{\varphi}\apprge2m_{\nu_{R}}$.
The mass of the canonically normalized field $\varphi$ at the minimum
of the potential is given by the second derivative of the potential
(\ref{varphipot}) 
\[
m_{\varphi}=\sqrt{V_{\varphi,\,\varphi}^{E}}\Big\vert_{\varphi=\langle\varphi\rangle}=2\times10^{-6}\mu,
\label{infmas}
\]
where we have taken a value for $A\sim5\times10^{-12}$ from Table.~\ref{tab1}.



\begin{figure}[H]
\centering\includegraphics[height=2.5in]{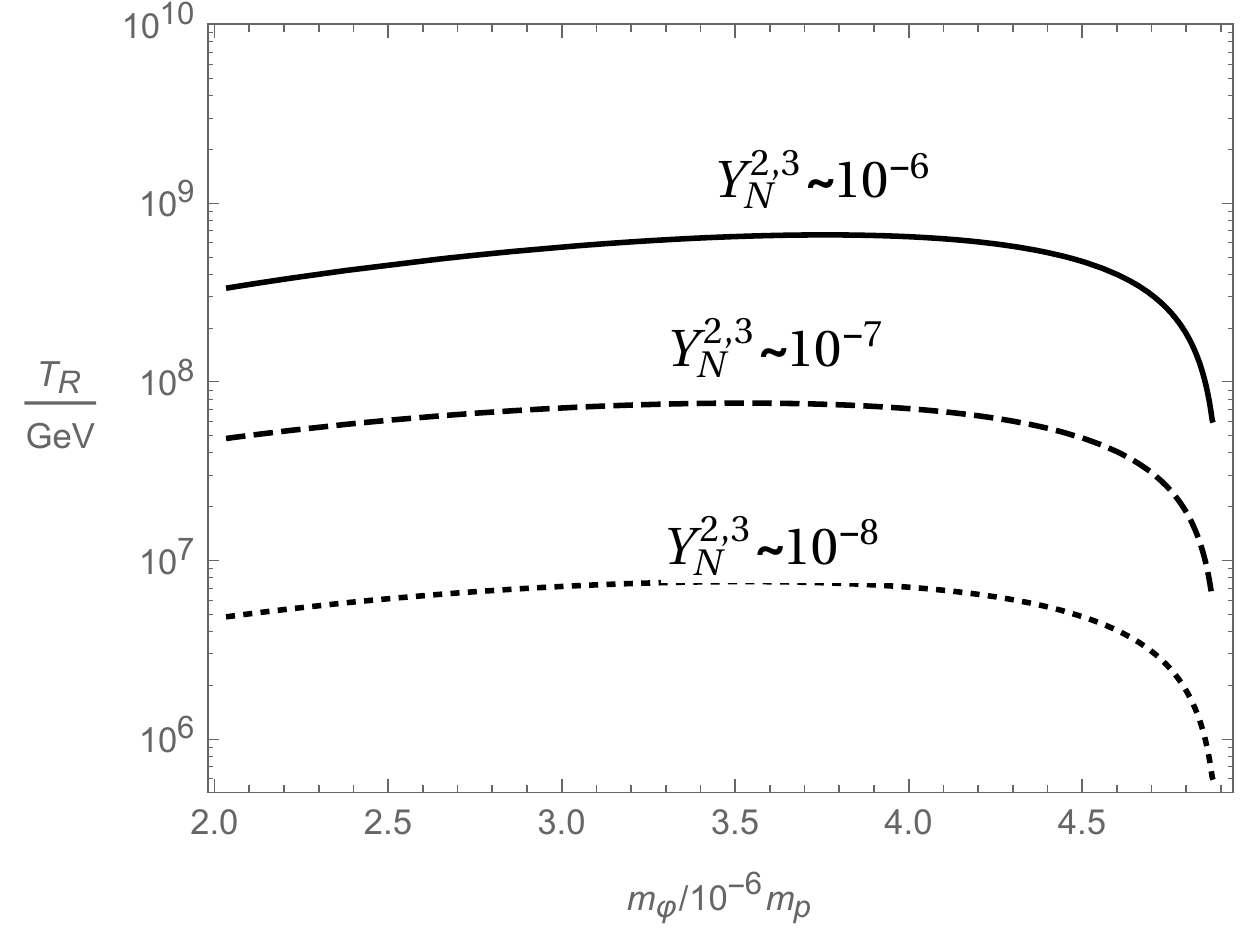}\caption{In this plot we depict the reheating temperatures $T_{R}$ Vs. $m_{\varphi}$
for the values of couplings $Y_{N}^{2,3}\sim10^{-8}-10^{-6}$.}

\label{TRm} 
\end{figure}

We implement the scheme of non-thermal leptogenesis proposed in \cite{Fukugita:1986hr,Lazarides:1991wu}
which can give rise to baryogenesis through CP violating decays
of RH Majorana neutrinos. In this section, we closely follow \cite{Asaka:2002zu,Senoguz:2003hc,Senoguz:2005bc}. We consider:
\begin{itemize}
\item Hierarchical masses for RHNs $m_{\nu_{R}^{1}}\ll m_{\nu_{R}^{2}}\sim m_{\nu_{R}^{3}}$.
To arrange this we require the coupling constants to be $Y_{N_{1}}\ll Y_{N_{2}}\sim Y_{N_{3}}$.
{We assume that the inflaton decays equally into the two
heavy RHNs $\nu_{R}^{2,3}$ and the corresponding
reheating temperature can be computed using }\cite{Asaka:2002zu,Okada:2013vxa}
\end{itemize}
\begin{equation}
T_{R}=\left(\frac{90}{\pi^{2}g_{*}}\right)^{1/4}\sqrt{\Gamma_{\varphi}\left(\varphi\to\nu_{R}^{i}\nu_{R}^{i}\right)m_{\rm P}}\,,\label{TRexp}
\end{equation}
where $g_{*}=105.6$ is the number of relativistic degrees of freedom
and the decay rate is given by 

\[
\Gamma_{\varphi}\left(\varphi\to\nu_{R}^{i}\nu_{R}^{i}\right)\simeq\frac{m_{\varphi}}{4\pi}\sum_{i=1}^{3}c_{i}^{2}\left(\frac{m_{\nu_{R}^{i}}}{m_{\rm P}}\right)^{2}\left(1-\frac{4m_{\nu_{R}^{i}}^{2}}{m_{\varphi}^{2}}\right)^{3/2}\,.
\]
{The masses of heavy RHNs are $m_{\nu_{R}^{2,3}}\sim\frac{Y_{N}^{2,3}}{\sqrt{2}}$,
which for $Y_{N}^{2,3}\sim10^{-8}-10^{-6}$ we have $m_{\nu_{R}^{2,3}}\sim10^{10}-10^{12}$
GeV. In Fig. \ref{TRm} we plot the possible reheating temperatures\footnote{Even though our model in this paper is non-SUSY, it is worth to mention herein the SUSY setup, where the reheating temperature is constrained by gravitino production and the corresponding leptogenesis \cite{Khlopov:1984pf,Falomkin:1984eu,Khlopov:1993ye,Khlopov:2004tn}}
of our case taking $c_{1}\approx0$ and $c_{2}=c_{3}=1$.}

\begin{itemize}
\item The decays of RH Majorana neutrinos $\nu_{R}^{i}$ break the lepton number
conservation and leads to CP violation. There are two decay channels
\end{itemize}
\begin{equation}
\Gamma_{i}:\nu_{R}^{i}\to H+l_{i}\,,\quad\bar{\Gamma}_{i}:\nu_{R}^{i}\to H^{\dagger}+\bar{l}_{i}\,,\label{channels-nuR}
\end{equation}
where $H$ and $l$ denote the Higgs field and the lepton doublets
of the SM. The (lepton asymmetry generated by the CP violation) decay
of $\nu_{R}^{i}$ is measured by the following quantity 

\begin{equation}
\epsilon_{i}\equiv\frac{\Gamma_{i}-\bar{\Gamma}_{i}}{\Gamma_{i}+\bar{\Gamma}_{i}}\lll1\,.\label{epsi}
\end{equation}
{CP asymmetry $\epsilon_{i}$ can be computed for
the dominant decays of $\nu_{R}^{2,3}$ using \cite{Flanz:1994yx,Buchmuller:1997yu,Hamaguchi:2002vc,Senoguz:2003hc}}

{
\begin{equation}
\epsilon_{i}=-\frac{1}{8\pi}\frac{1}{\left(Y_{D}Y_{D}^{\dagger}\right)_{11}}\sum_{i=2,3}\text{Im}\left[\left\{ \left(Y_{D}Y_{D}^{\dagger}\right)_{1i}\right\} ^{2}\right]\left[f\left(\frac{m_{\nu_{R}^{i}}^{2}}{m_{\nu_{R}^{1}}^{2}}\right)+g\left(\frac{m_{\nu_{R}^{i}}^{2}}{m_{\nu_{R}^{1}}^{2}}\right)\right]\,,\label{epsilon_i-exp}
\end{equation}
where
\[
f\left(y\right)=\sqrt{y}\left[-1+\left(y+1\right)\ln\left(1+\frac{1}{y}\right)\right]\,,\quad g\left(y\right)=\frac{\sqrt{y}}{y-1}\,.
\]
Here, we only aim to constrain the range of values for $\epsilon_{i}$
leaving for future studies the explicit computation of constraining Yukawa matrix $Y_{D}^{ij}$ \cite{Asaka:2002zu}. }

The lepton asymmetry is given by

\begin{equation}
\frac{n_{L}}{s}=\sum_{i=1}^{3}\epsilon_{i}\text{Br}_{i}\frac{3T_{R}}{2m_{\varphi}}\,,\label{nLs}
\end{equation}
where $n_{L}$ is the difference between number of leptons and anti-leptons
and $s$ indicates the entropy density, $\text{Br}_{i}$ denotes the
branching ratio 
\begin{itemize}
\item The production of RH Majorana neutrinos happens non-thermally and sufficiently
late so that the produced lepton asymmetry sources the baryon asymmetry
at a later stage. This essentially requires $m_{\nu_{R}^{1}}\gtrsim T_{R}$
so that the later decay of lightest RH Majorana neutrino $\nu_{R}^{1}$
does not wash away the produced lepton asymmetry by the heavy ones.
We assume there is an accidental $B-L$ conservation\footnote{$B,\,L$ refers to baryon number and lepton number, respectively. }
such that sphaleron process is active which brings a part of the above
lepton asymmetry into the baryon asymmetry (see Ref. \cite{Khlebnikov:1988sr,Harvey:1981yk,Harvey:1990qw}
for details). {As the reheating temperature in our
case is $T_{R}\sim10^{6}-10^{9}\,\text{GeV}$ (see Fig. \ref{TRm}), we take
$Y_{N}^{1}\sim10^{-10}-10^{-9}$ such that $m_{\nu_{R}^{1}}\sim10^{8}-10^{9}\,\text{GeV}$
. Therefore, with values $m_{\nu_{R}^{2,3}}\sim10^{10}-10^{12}\,\text{GeV}$
, $m_{\nu_{R}^{1}}\sim10^{8}-10^{9}\,\text{GeV}$ and $T_{R}\sim10^{6}-10^{9}\,\text{GeV}$,
we have met the conditions for successful leptogenesis which are $m_{\nu_{R}^{2}}\sim m_{\nu_{R}^{3}}\gg m_{\nu_{R}^{1}}$
and $m_{\nu_{R}^{1}}\gtrsim T_{R}$.}
\end{itemize}
Baryon asymmetry is proportional to the lepton asymmetry as 

\begin{equation}
\begin{aligned}\frac{n_{B}}{s}\simeq & \frac{28}{79}\frac{n_{L}}{s}\\
\simeq & \frac{42}{79}\sum_{i=1}^{3}\epsilon_{i}\text{Br}_{i}\frac{T_{R}}{m_{\varphi}}\,.
\end{aligned}
\label{nBsnls}
\end{equation}

The baryon asymmetry which is measured by the ratio of the difference
between the number of baryons minus the anti-baryons $n_{B}$ to the
entropy density in the present Universe is constrained \cite{Ade:2015xua} in the following form

\begin{equation}
\frac{n_{B}}{s}=\left(6.05\pm0.06\right)\times10^{-10}\,.\label{nBs}
\end{equation}
Considering branching ratios $\text{Br}_{1}=0$ and $\text{Br}_{2}=\text{Br}_{3}=\frac{1}{2}$
with $\epsilon_{1}\ll\epsilon_{2}\sim\epsilon_{3}$, we have 

\begin{equation}
\frac{n_{B}}{s}\approx\epsilon_{2}\frac{T_{R}}{m_{\varphi}}\,.\label{nbsapp}
\end{equation}
{From Fig. \ref{TRm} we can read that $\frac{T_{R}}{m_{\varphi}}\sim10^{-7}\sim10^{-4}$
, which indicates the CP violation in the decay of RH Majorana neutrinos
$\left(\epsilon_{i}\right)$ must be in the range $6\times10^{-6}\lesssim\epsilon_{2,3}\lesssim6\times10^{-3}$
to have the observed baryon asymmetry.} 

\section{Conclusions}

\label{conclusions}

Coleman-Weinberg inflation \cite{Shafi:1983bd} has been a successful
and realistic model based on GUT and is consistent with the current Planck
data with $r\gtrsim0.02$ \cite{Okada:2014lxa}. In this work, we
have further generalized the framework of CW inflation with an additional
conformal symmetry. Spontaneous breaking of conformal symmetry is useful to 
create a hierarchy of mass scales, therefore it is natural to realize this symmetry in GUT models. 
In this respect, two complex singlet fields of
$\text{SU}(5)$ or $SO(10)$ were considered and are coupled to the GUT fields in
a suitable manner. We have showed that this setup, upon spontaneous
breaking of GUT and conformal symmetry, leads to an interesting inflationary
scenario driven by the real part of the singlet field. In our model,
the above VEV branch of CW potential gets flattened to a Starobinsky
plateau, allowing for $n_{s}\sim 1-\frac{2}{N}$ and $r\sim \frac{12}{N^2}$
for $N\sim50-60$ number of e-foldings. {Therefore, our model is observationally fits with the same predictions of the Starobinsky and Higgs inflation.} 
Moreover, the VEV of the inflaton affects the masses of the superheavy gauge bosons
that mediate the proton decay. We calculated the corresponding estimates
for the proton life time above the current lower bound from Super-K
data as $\tau_{p}\left(p\to\pi^{0}+e^{+}\right)>1.6\times10^{34}$. In
the next step, we introduced a coupling between the complex singlet field
with the generation of three singlet RHNs, where the singlet
field is assumed to carry two units of lepton number. We implemented
a type I seesaw mechanism, where spontaneous symmetry breaking of global
lepton number results in generating neutrino masses. We put an upper
bound to the inflaton couplings to RHNs, assuming inflation is dominated by inflaton 
couplings to GUT field. For the non-thermal
leptogenesis to happen, we have considered a dominant decay of the inflaton
into some of the RHNs and obtained the corresponding reheating
temperatures as $10^{6}\text{ GeV}\lesssim T_{R}<10^{9}$ GeV. Furthermore, our proposed extension of CW inflation can be tested within
future CMB and particle physics experiments \cite{Creminelli:2015oda}. 

In this work, we mainly restricted to a non-supersymmetric construction of GUT inflation with conformal symmetry. It would be interesting to consider this model in GUT based SUGRA framework with superconformal symmetries, which we defer for
future investigations. 

\acknowledgments

We thank the anonymous referee for very useful comments. 
We would like to thank Qaisar Shafi for numerous useful discussions and feedback during this project. 
We would like to thank C.~Pallis, N.~Okada and D.~Raut for useful discussions and comments. K.~S.~K thanks P. Parada and K.~N.~Deepthi for useful discussions. 
This research work was supported by the grant UID/MAT/00212/2013 and
COST Action CA15117 (CANTATA). K.~S.~K thanks
FCT BD grant SFRH/BD/51980/2012 and  Netherlands Organization for Scientific Research (NWO) grant number 680-91-119.  . PVM is grateful to DAMTP, University of Cambridge for providing an excellent research environment for his sabbatical and he is also thankful to Clare Hall college,  Cambridge for a Visiting Fellowship.

\appendix

\section{Geometry of fields space}

\label{A1}

In the action (\ref{CFTSU(5)}) we have primarily three fields $\LF \phi=\sqrt{2}\mathfrak{Re}\left[\Phi\right],\, \tau=\text{Im}\LT \Phi\RT,\,\sigma\RF$ which are non-minimally coupled to the Ricci scalar\footnote{Note that we gauge fixed the conformal field at $X=\sqrt{3}M$}. After a conformal transformation to the Einstein frame, all their kinetic terms get in general modified, therefore introducing a field space geometry \cite{Kaiser:2010ps}. In more detail, rewriting our action (\ref{CFTSU(5)})  in terms of the  three fields $\LF \phi=\sqrt{2}\mathfrak{Re}\left[\Phi\right],\, \text{Im}\LT \Phi,\RT, \sigma\RF$, we retrieve 

\begin{equation}
S_G= \int d^4x\sqrt{-g}\LT \LF 6M^2-\phi^2- \sigma^2-\tau^2\RF \frac{R}{12}-\frac{1}{2}\partial^\mu\phi\partial_\mu \phi-\frac{1}{2}\partial^\mu \sigma \partial_\mu\sigma-\frac{1}{2}\partial^\mu\tau\partial_\mu\tau -V\LF \phi,\,\tau,\,\sigma \RF \RT \,, 
\label{EA1}
\end{equation}
where $V\LF \phi,\,\tau,\,\sigma \RF  $ is the potential whose details are not relevant here. A conformal transformation of the action (\ref{EA1}) into the Einstein frame yields (expressing in the units of $m_P=1$)

\begin{equation}
S_E= \int d^4x \sqrt{-g} \LT \frac{1}{2}R_E- \frac{1}{2}G_{IJ} \partial^\mu\phi^I\partial_\mu\phi^J-\frac{V\LF \phi,\,\tau\,,\sigma \RF}{\Omega_1^2} \RT\,,
\end{equation}
where $\Omega_1 = \LF 6M^2-\phi^2- \sigma^2-\tau^2\RF$ and $G_{IJ}$ is the fields space metric  

\begin{equation}
G_{IJ}= \frac{M^2}{2\Omega_1}\delta_{IJ}+ \frac{3}{2}\frac{M^2}{\Omega_1^2}\Omega_{1,I}\Omega_{1,J}\,,
\label{fieldmet}
\end{equation}
where $\Omega_{1,I}= \frac{\partial\Omega_1}{\partial\phi^I}$. 
The field metric (\ref{fieldmet}) provides the dynamics of the fields in the Einstein frame. Here we label $I, J= 1,\,2,\,3$ as $\phi, \tau, \sigma$, respectively. 
It was shown in  \cite{Kaiser:2010ps,Renaux-Petel:2015mga} that if the scalar curvature of fields space is negative it might lead to geometrical destabilization during or after inflation depending on the dynamics of inflaton and the heavy fields.

To make the analysis easy let us first analyse the two fields space $\LF \phi,\,\tau \RF$ in the direction of 

\begin{equation}
\sigma^2 = \frac{2\lambda_{2}}{\lambda_c} \phi^2 f\LF \frac{\phi}{\sqrt{6}M} \RF\,. 
\end{equation}
Let us consider $f\LF \frac{\phi}{\sqrt{6}} \RF = \LF 1-\frac{\phi^2}{6M^2} \RF$. As discussed in Sec.~\ref{twofieldmodel-sec} during the inflationary regime $\phi \to \sqrt{6}M$ and since we have assumed $\lambda_{2} \ll \lambda_c$, then we have $\sigma^2\ll \phi^2$. In this phase, action (\ref{EA1})  effectively reduces to 

\begin{equation}
S_G \approx \int d^4x \sqrt{-g} \LT \LF 6M^2-\phi^2-\tau^2 \RF \frac{R}{12}-\frac{1}{2}\partial_\mu\tau\partial^\mu\tau-\frac{1}{2}\partial^\mu\phi\partial_\mu\phi - V\LF \phi,\,\tau \RF \RT \,.
\label{EA2}
\end{equation}
We can re-parametrize the fields as
\begin{equation}
\phi = \rho\sin\theta\, \quad \tau = \rho \cos\theta \, , 
\label{param}
\end{equation}
where $\theta = \tan^{-1}\LF \frac{\phi}{\tau} \RF$. Note that inflaton trajectory considered in Sec.~\ref{twofieldmodel-sec} corresponds to $\theta=0$. 
Substituting (\ref{param}) in the action (\ref{EA2}) we get 
\begin{equation}
S_G \approx \int d^4x\sqrt{-g}\LT \LF 6M^2-\rho^2 \RF-\frac{1}{2}\partial_{\mu}\rho\partial^\mu\rho-\frac{1}{2}\rho^2\partial^\mu\theta\partial_\mu\theta  - V\LF \rho,\,\theta \RF \RT\,. 
\end{equation}
Conformally transforming the above action into the Einstein frame gives  (in the units of $m_P=1$)
\begin{equation}
S_G \approx \int d^4x\sqrt{-g}\LT \frac{1}{2}R_E-\frac{1}{2M^2\LF 1-\frac{\rho^2}{6M^2}\RF^2}\partial_{\mu}\rho\partial^\mu\rho-\frac{\rho^2}{2\LF 6M^2-\rho^2\RF}\partial^\mu\theta\partial_\mu\theta - \frac{V\LF \rho,\,\theta\RF}{\LF 6M^2-\rho^2 \RF^2} \RT\,. 
\end{equation}
The above action can be rewritten as the following, by introducing a field metric 

\begin{equation}
S_G \approx \int d^4x\sqrt{-g}\LT \frac{1}{2}R_E-\frac{1}{2}G_{IJ}\partial^\mu\phi^I\partial_\mu\phi^J - \frac{V\LF \rho,\,\theta\RF}{\LF 6M^2-\rho^2 \RF^2} \RT\,, 
\end{equation}
where 
\begin{equation}
G_{IJ}=\left(\begin{array}{cc}
\frac{36M^{2}}{\left(6M^{2}-\rho^{2}\right)^{2}} & 0\\
0 & \frac{1}{\left(6M^{2}-\rho^{2}\right)}
\end{array}\right)\,.
\label{fiemet}
\end{equation}
Here $I,\,J=1,2$ for $\rho,\,\theta$ respectively. Computing the Ricci tensor and Ricci scalar for the metric  (\ref{fiemet}) we obtain 

\begin{equation}
\mathcal{R}_{IJ}= \left(
\begin{array}{cc}
-\frac{1}{2} & 0 \\
0 & -\frac{1}{2} \sinh ^2\left(\frac{\rho }{\sqrt{2}}\right)\,. 
\end{array}
\right)\,,\quad \mathcal{R}=-1\,. 
\end{equation}
Notice that the Ricci scalar associated to the fields space is negative and unit.  It is very similar to several Starobinsky like models of inflation and $\alpha-$ attractor models of SUGRA\footnote{The Ricci scalar from a fields space metric in the case of $\alpha-$ attractor models is $\mathcal{R}_K=-\frac{2}{3\alpha}$ \cite{Renaux-Petel:2015mga}.}, for which it was shown that geometrical destabilization could only occur towards the end of inflation \cite{Renaux-Petel:2015mga}. The point to emphasize the following. With a suitable choice of potential for $\theta$, fields space geometrical effects on inflationary epoch can be heavily suppressed (see e.g., \cite{Achucarro:2017ing} for more details). However, in recent studies, effects of heavy fields during preheating epoch have been explored in multifield non-canonical, non-minimal models of inflation \cite{Ema:2016dny,DeCross:2016fdz,Krajewski:2018moi,Iarygina:2018kee,Christodoulidis:2019mkj}. We opted to ignore such effects in our investigation and assume likewise that inflaton dominantly decays into the RHNs in Sec.~\ref{ReheatSec}. We leave for future an analysis of this interesting aspect. 

Let us now consider the two fields space $(\phi,\,\sigma)$ in the direction of $\text{Im}[\Phi]=0$ (i.e., $\theta=0$) 

\begin{equation}
S_G = \int d^4x\sqrt{-g}\LT \LF 6M^2-\phi^2-\tilde{\sigma}^2\RF \frac{R}{12} - \frac{1}{2}\partial_\mu\phi\partial^\mu\phi -\frac{1}{2}\partial_\mu\tilde{\sigma}\partial^\mu\tilde{\sigma} -V\LF \phi,\,\tilde{\sigma} \RF\RT\,,
\end{equation}
where we have rescaled $\sigma\to \frac{\sigma}{\sqrt{2}}$.
Computing the Ricci scalar associated to fields space $(\phi,\,\tilde{\sigma})$ we obtain \cite{Kaiser:2010ps}
\begin{equation}
\tilde{\mathcal{R}} = \frac{72M^2}{\LT 2\omega+6\LF \omega_{,\phi}^2+\omega_{,\tilde{\sigma}}^2 \RF  \RT^2}\,, 
\label{riccif2}
\end{equation}
where $\omega= 6M^2-\phi^2-\tilde{\sigma}^2$ and $\omega_{,\phi}= \frac{\partial\omega}{\partial\phi},\,\omega_{,\tilde{\sigma}}= \frac{\partial\omega}{\partial\tilde{\sigma}}  $. We can clearly see that the Ricci scalar of fields space $\LF \phi,\,\tilde{\sigma} \RF$ is positive and therefore geometric destabilization may not occur in this case during inflationary regime. Even though efficient particle production can occur in the models with a positive field space with sharp features \cite{Ema:2016dny,DeCross:2016cbs,DeCross:2016fdz,DeCross:2015uza,Nguyen:2019kbm}. 

\bibliographystyle{utphys}
\bibliography{References}

\end{document}